\documentclass[aps,prb,twocolumn,showpacs,amsmath,amssymb,superscriptaddress,floatfix]{revtex4-2}
\usepackage[english]{babel}
\usepackage{blindtext}
\usepackage{latexsym}
\usepackage{amssymb}
\usepackage{physics}
\usepackage{amsmath}
\usepackage{bm}
\usepackage{mathtools}
\usepackage{amsfonts}
\usepackage{relsize}
\usepackage{xcolor}
\usepackage{verbatim}
\usepackage{bbold}
\usepackage{slashed}
\usepackage{appendix} 
\usepackage{graphicx}
\usepackage[normalem]{ulem}
\usepackage[colorlinks = true,
            linkcolor = blue,
            urlcolor  = blue,
            citecolor = blue,
            anchorcolor = blue]{hyperref}
\usepackage{graphicx} 
\usepackage{dcolumn} 
\newcommand{\approptoinn}[2]{\mathrel{\vcenter{
  \offinterlineskip\halign{\hfil$##$\cr
    #1\propto\cr\noalign{\kern2pt}#1\sim\cr\noalign{\kern-2pt}}}}}

\newcommand*{\colorboxed}{}
\def\colorboxed#1#{%
  \colorboxedAux{#1}%
}
\newcommand*{\colorboxedAux}[3]{%
  \begingroup
    \colorlet{cb@saved}{.}%
    \color#1{#2}%
    \boxed{%
      \color{cb@saved}%
      #3%
    }%
  \endgroup
}

\newcommand{\bs}[1]{\mathbf{#1}}

\newcommand{\bq}{\bs{q}}
\newcommand{\bQ}{\bs{Q}}

\newcommand{\bk}{\bs{k}}

\newcommand{\eps}{\epsilon}

\definecolor{ao(english)}{rgb}{0.0, 0.5, 0.0}
\definecolor{amaranth}{rgb}{0.9, 0.17, 0.31}
\definecolor{green(html/cssgreen)}{rgb}{0.0, 0.5, 0.0}

\begin{document}
\title{Mean-field analysis of a Hubbard interaction on Bernal Bilayer Graphene}
\author{Robin Scholle}
\affiliation{Max Planck Institute for Solid State Research, D-70569 Stuttgart, Germany}
\author{Laura Classen}
\affiliation{Max Planck Institute for Solid State Research, D-70569 Stuttgart, Germany}
\affiliation{School of Natural Sciences, Technische Universität München, D-85748 Garching, Germany}
\date{\today}
\begin{abstract}
We perform unrestricted Hartree-Fock calculations on the 2D Hubbard model on a 
honeycomb and bilayer honeycomb lattice 
at both zero and finite temperatures. Finite size real space calculations are supplemented with RPA calculations in the thermodynamic limit. Our motivation comes from high doping levels achieved in graphene and Bernal bilayer graphene by interacalation. We present phase diagrams in doping and temperature for a moderate Hubbard interaction. The magnetic states we find are classified systematically based on the dominant Fourier components of their spin patterns, their average magnetization and spin incommensurabilities. The dominant spin patterns are N\'eel order and various types of stripes. Around Van Hove filling, we resolve the competition between stripe and chiral spin density waves in the symmetry-broken regime.  
We also investigate the effect of an applied external displacement field on the spin patterns of BBG. 
\end{abstract}
\maketitle


\section{Introduction}

The Hubbard model in its minimalist setup is a key model for understanding strongly correlated materials. Despite its simplicity, it can give rise to a rich variety of interaction-induced ordered phases and more exotic correlation phenomena, such as antiferromagnetism, superconductivity, or a pseudogap phase. While impressive progress on its phase diagram was achieved, the large possible parameter space continues to provide unexplored regions, including its application to different lattice geometries in two spatial dimensions \cite{Arovas2022,Qin2022}. 

In the most prominent representative - the square-lattice Hubbard model - a Ne\'el antiferromagnet arises at half filling, and makes room for various magnetic or charge-ordered phases, such as spirals \cite{Shraiman1989, Shraiman1992, Chubukov1992, Chubukov1995, Dombre1990, Fresard1991, Kotov2004, Igoshev2010, Igoshev2015, Yamase2016, Eberlein2016, Mitscherling2018, Bonetti2020a} or stripes \cite{Schulz1989, Zaanen1989, Machida1989, Poilblanc1989, Schulz1990, Kato1990, Seibold1998, Fleck2000, Fleck2001, Raczkowski2010, Timirgazin2012, Peters2014, Matsuyama2022, Zheng2017, Qin2020}. 
Due to the many possible states, an unbiased approach towards this zoo of states is desirable and extensive numerical efforts start to converge towards a phase diagram \cite{PhysRevX.11.011058,Qin2022}. Interestingly, many of these ordered phases already appear within mean-field theory, and a recent study revealed the Ne\'el-to-spiral-to-stripe transitions using unrestricted Hartree Fock calculations in real space \cite{Scholle2023,ScholleSpiralToStripe}. While such mean-field calculations tend to overestimate ordering tendencies and violate the Mermin-Wagner theorem, they still provide valuable insights about dominant correlations and can serve as guidance for numerical studies or gauge theories of fluctuating magnetic order \cite{Scheurer2018, Sachdev2019review, Bonetti2022gauge}. 

It is an open question how much of the square-lattice phenomenology can be transferred to other lattices. We take this as a motivation to study the phase diagram of the Hubbard model on a single- and bi-layer honeycomb lattice. The model can be used to study interaction effects in graphene and Bernal bilayer graphene (BBG) \cite{RevModPhys.81.109,McCann_2013}. 
Previously, interaction-induced ordered states in these systems were theoretically studied mainly at charge neutrality and Van Hove filling. A competition between different quantum spin, charge, or valley Hall states arises at charge neutrality, which a (layered) Ne\'el antiferromagnet was predicted to win for a Hubbard interaction above a critical value \cite{PhysRevLett.100.146404,PhysRevB.85.235408,PhysRevB.81.041401,PhysRevB.86.115447,PhysRevB.86.075467,PhysRevB.73.214418,PhysRevB.82.115124,PhysRevB.103.205135,Meng2010,PhysRevLett.109.126402,ZHANG20159,PhysRevB.101.125103,PhysRevB.81.115416,S.Sorella_1992}. To see this in BBG, trigonal warping or self-energy effects need to be taken into account \cite{PhysRevLett.117.086404}. 
Around Van Hove filling, the interplay between magnetism and superconductivity was investigated \cite{nandkishore2012chiral,PhysRevB.86.115426,PhysRevB.89.144501,PhysRevB.86.020507,PhysRevB.102.125141}. An approximately nested Fermi surface at this filling enables three degenerate spin density wave states with symmetry-related nesting vectors $M_i$. Their collinear or orthogonal combination leads to a gapped chiral or a half-metallic uniaxial spin density wave \cite{Li_2012,PhysRevLett.101.156402,Nandkishore2012,PhysRevLett.105.266405}. 

Experimentally, single-layer graphene is a stable semimetal at charge neutrality. In contrast, suspended samples of BBG show signatures of interaction-induced ordered states including a layered antiferromagnet, quantum anomalous Hall, and nematic states \cite{PhysRevLett.106.156801,doi:10.1126/science.1194988,Velasco2012,PhysRevLett.108.076602,PhysRevLett.105.256806,Ki2013,doi:10.1126/science.1208683,Feldman2009,Geisenhof2021}. Furthermore, it was demonstrated that high doping regimes can be achieved by intercalation in both single- and bi-layer graphene \cite{Yang_2022,PhysRevB.100.121407,PhysRevLett.125.176403,PhDLink,PhysRevLett.104.136803}. Single-layer graphene was even doped beyond its Van Hove singularity \cite{PhysRevLett.125.176403}. This promises the possibility to also realise interaction-induced ground states in highly doped graphene systems. 

In this work, we perform a comprehensive analysis of the phase diagram of graphene and BBG as modeled by a Hubbard model on the single- or bi-layer honeycomb lattice. We employ unrestricted Hartree Fock calculations in real space, complemented by an instability analysis via a random phase approximation (RPA) in the thermodynamic limit. We complete the phase diagram for the entire range of doping from empty to fully filled bands at zero and finite temperatures. 
Similar to the square-lattice Hubbard model, we find that a (layered) Ne\'el antiferromagnet transitions into a cascade of stripe phases. However, for larger dopings the phase diagram on the single- and bi-layer honeycomb lattices is qualitatively different due to their Van Hove singularities. We observe a competition between spin density waves with three Fourier modes. This includes the detailed temperature and density dependence of the chiral and uniaxial spin density waves which were previously studied at Van Hove doping. 
Furthermore, we analyse the effect of unequal layer occupation in BBG, which can arise from asymmetric intercalation or an external displacement field. We note that we focus on density regimes beyond the cascade behavior recently observed in BBG in a displacement field at the lowest densities \cite{doi:10.1126/science.abm8386,Seiler2022,PhysRevLett.133.066301}. We find that for intermediate interaction strengths, the overall field dependence can be well understood based on the phase diagram of two almost decoupled single layers. 

The rest of the paper is structured as follows. In Sec.~\ref{sec:model}, we introduce the Hubbard model on a single-layer and Bernal-stacked bilayer honeycomb lattice, and we describe our unrestricted Hartree Fock and RPA calculations. Furthermore, we provide a classification of the ordered states we find based on their real-space spin and charge patterns, and Fourier modes. In Sec.~\ref{sec:results}, we present the phase diagram, magnetisation, and wave vectors of the ordered states for the single- and bi-layer case. We vary doping, temperature, and for BBG a displacment field. We provide a discussion of our results and a brief outlook about possible further research directions in Sec.~\ref{sec:conclude}.

\section{Model and Method}
\label{sec:model}

\begin{figure}
    \centering
    \includegraphics[width=0.4\textwidth]{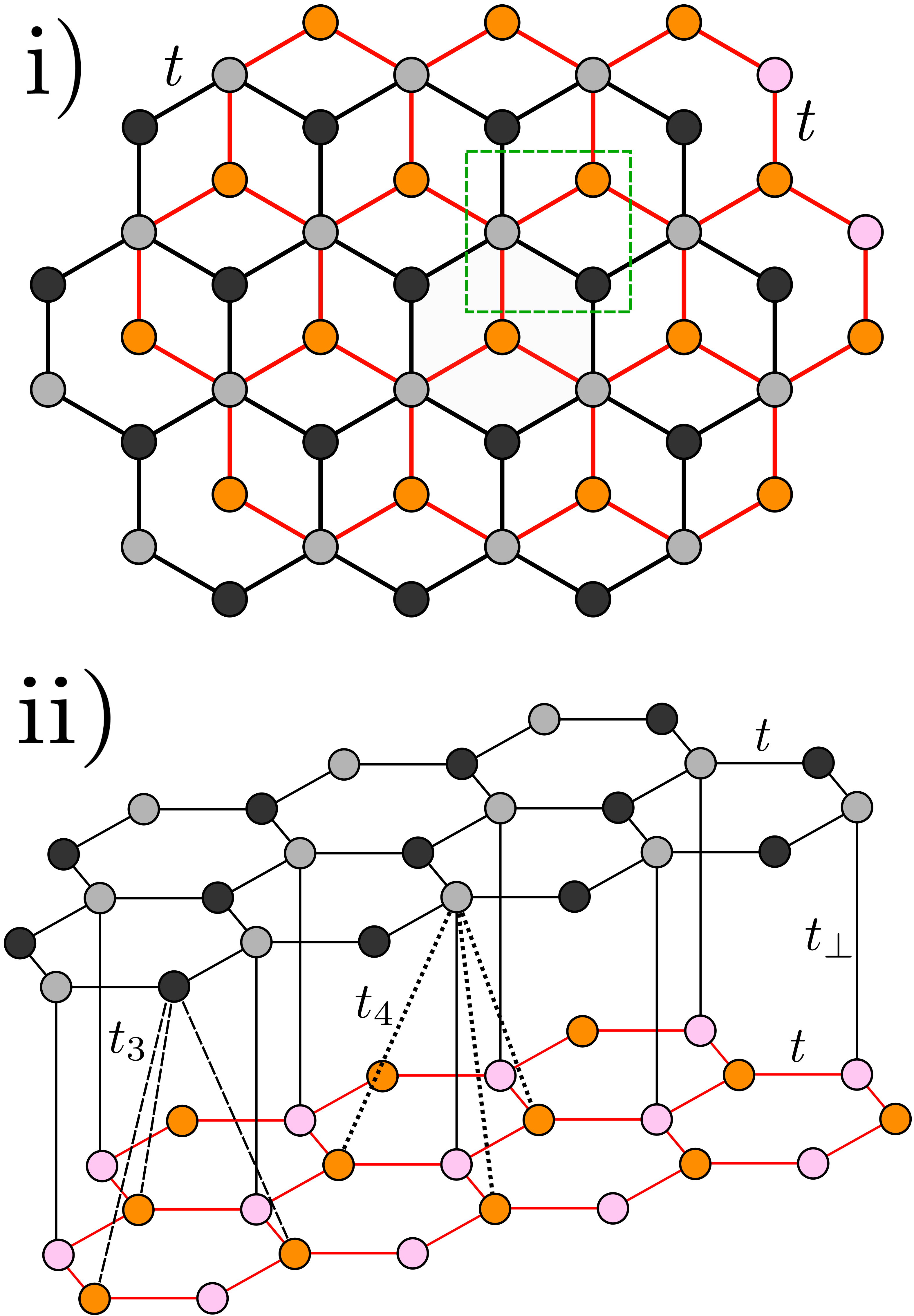}
    \caption{Overview of the real-space structure of Bernal bilayer graphene  
    showing  
    i) the top view 
    and ii)  
    a view from the side with an indication of the four different types of hoppings, $t, t_\perp, t_3$ and  $t_4$. The green dashed box is one possible choice for a BBG unit cell, which contains 4 sites.}
    \label{fig: Skizze BBG}
\end{figure}

Motivated by graphene and Bernal bilayer graphene, we consider the Hubbard model on a single- and AB-stacked bi-layer honeycomb 
lattice  
\begin{equation} \label{eq: HubbardHamiltonian}
 H = H_0 + H_{\text{int}} =
 \sum_{j,j',\sigma} t_{jj'} c^\dagger_{j\sigma} c^{\phantom\dagger}_{j'\sigma} +
 U \sum_j n_{j\uparrow} n_{j\downarrow} , 
\end{equation}
with local interaction $U>0$ and hopping amplitudes $t_{jj^\prime}$. Here, 
$c^\dagger_{j\sigma}$ ($c_{j\sigma}$) is the fermion creation (annihilation) operator for an electron with spin $\sigma$ on site $j$, and $n_{j\sigma} = c^\dagger_{j\sigma} c^{\phantom\dagger}_{j\sigma}$. The index $j$ runs over all atoms in both layers. The hopping matrix $t_{jj'}$ depends only on the distance of the sites $j$ and $j^\prime$. 
We model the single-layer case with only the nearest-neighbor hopping amplitude $t$ being non-zero. In the bilayer case, we include nearest-neighbor hopping within a layer, and vertical $t_\perp = 0.121t$ and skew hopping terms
$t_3 = 0.120t$, and $t_4 = 0.044t$ between layers \cite{McCann_2013}, see  
Fig. \ref{fig: Skizze BBG}.

\begin{figure}
    \centering
    \includegraphics[width=0.49\textwidth]{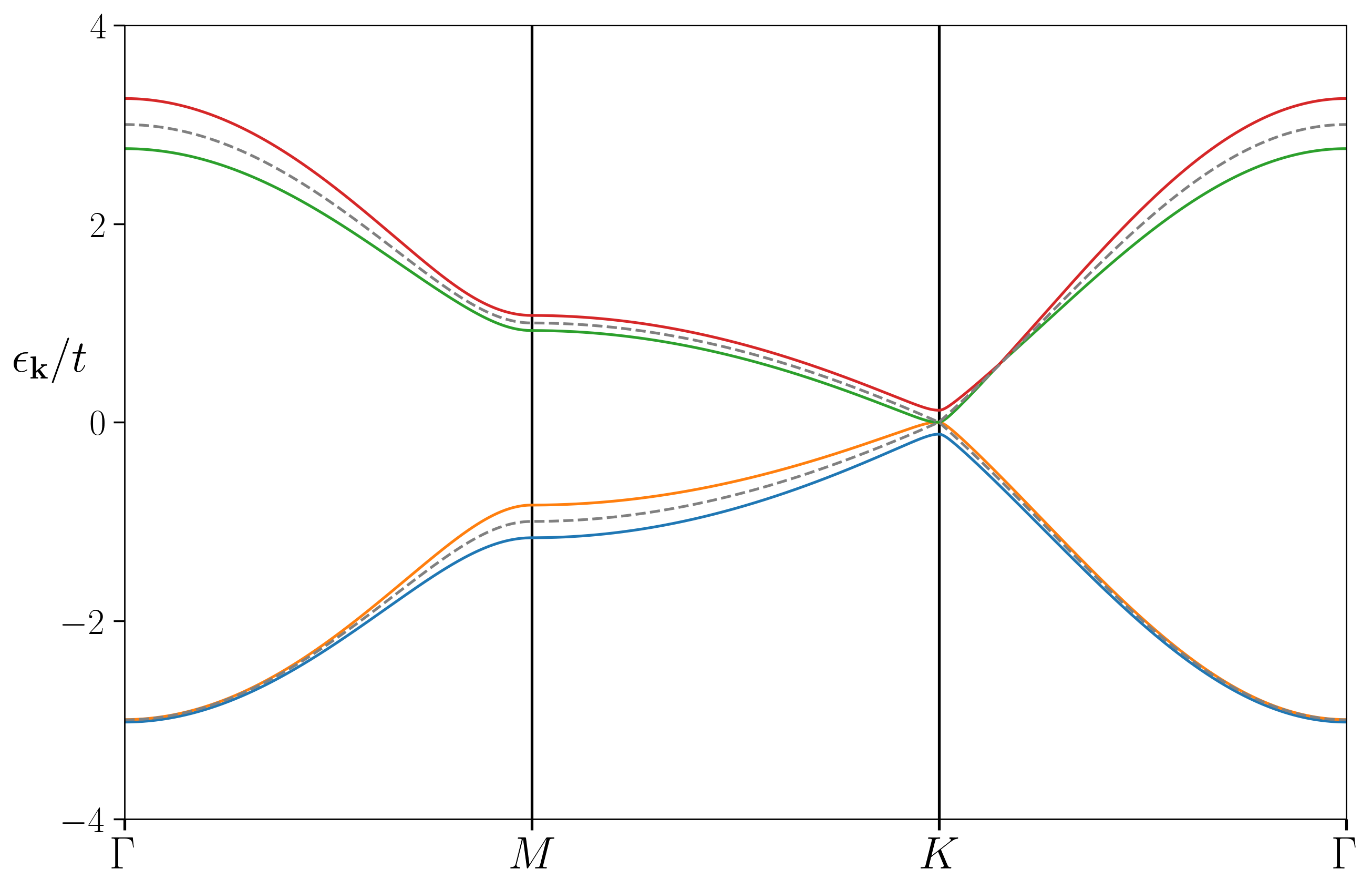}
    \caption{The bandstructure of BBG (colored solid lines) and single layer graphene (grey dashed lines). The bands are shown along the high symmetry path $\Gamma - M - K - \Gamma$.}
    \label{fig: BBG bandstructure}
\end{figure}

The unit cell of single-layer graphene contains two atoms (A,B) and that of bilayer graphene four (A${}_1$,B${}_1$,A${}_2$,B${}_2$) (see the green dashed box in Fig.~\ref{fig: Skizze BBG}i). In the sublattice basis and transforming to momentum space, the non-interacting Hamiltonian becomes $H_0=\sum_{\bk,\sigma}c^\dagger_{\bk\sigma}H_{0,\bk}c_{\bk\sigma}$, where $c_{\bk\sigma}=(c_{A\bk\sigma},c_{B\bk\sigma})^T$ or $c_{\bk\sigma}=(c_{A_1\bk\sigma},c_{B_1\bk\sigma},c_{A_2\bk\sigma},c_{B_2\bk\sigma})^T$, and for single-layer graphene
\begin{equation}
\begin{split}
        H_{0,\bk} &=\left(\begin{array}{cc}
0 & -t\delta_\bk \\
-t\delta_\bk^* & 0 
\end{array}\right)\,,
\end{split}
\end{equation}
while for Bernal bilayer graphene
\begin{equation}
\begin{split}
        H_{0,\bk} &=\left(\begin{array}{cccc}
0 & -t\delta_\bk & t_4 \delta_\bk & -t_3 \delta_\bk^*	\\
-t\delta_\bk^* & 0 & t_\perp & t_4 \delta_\bk	\\
t_4 \delta_\bk^* & t_\perp & 0 & -t\delta_\bk	\\
-t_3 \delta_\bk & t_4 \delta_\bk^* &-t\delta_\bk^* & 0
\end{array}\right)
\end{split}
\end{equation}

with $\delta_\bk = e^{\frac{i k_y}{\sqrt{3}}} + 2 e^{\frac{-i k_y}{2\sqrt{3}}}\cos(k_x/2)$. We show the corresponding energy bands $\epsilon_\bk$ in Fig \ref{fig: BBG bandstructure}. They feature linear band touchings at half filling at the $\pm K$ 
points (see Fig.~\ref{fig: FourierWeights} for high-symmetry points in the Brillouin zone). BBG has 
three further linear band touching points very close to the $\pm K$ points. For large enough dopings of 25\%, a Van Hove singularity can be reached in single layer graphene. In bilayer graphene, there are two subsequent Van Hove energies of the two conduction/valence bands. The Fermi surface at the Van Hove energy is approximately nested with wave vector $M$.

In addition, the band structure of bilayer graphene can be tuned by an electric displacement field. For a non-zero displacement field, we add the term 

\begin{equation}
\label{eq:D}
    H_D = \frac{D}{2} \sum_j \epsilon_j n_j ,
\end{equation}
to the Hamiltonian where $\epsilon_j = 1$ in the upper layer and $\epsilon_j = -1$ in the lower layer. We note that doping of bilayer graphene via intercalation also leads to an asymmetry in the charge carrier density of both layers and can be modeled in an analogous way \cite{Yang_2022,PhDLink}.

\subsection{Unrestricted Hartree-Fock in real space} 

For an unbiased analysis of spin and charge orders, we perform a mean-field decoupling of the interaction term of Eq.~\eqref{eq: HubbardHamiltonian} in real space. The resulting mean-field expression is of the form  
\cite{Zaanen1989,Scholle2023}
\begin{eqnarray} \label{eq: InteractionTermHubbardModelMeanField}
 H_{\text{int}}^\mathrm{MF} &=&
 \sum_{j\sigma} \Delta_{j\overline{\sigma}} n_{j\sigma}
  + \sum_j \left( \Delta_{j-} c^{\dagger}_{j\uparrow} c_{j\downarrow}
    + \Delta_{j+} c^\dagger_{j\downarrow} c_{j\uparrow}\right) \nonumber \\
 && -\frac{1}{U}\sum_j \left( \Delta_{j\uparrow} \Delta_{j\downarrow}
   - \Delta_{j-} \Delta_{j+} \right) ,
\end{eqnarray}
where we defined $\overline{\uparrow} = \downarrow$ and $\overline{\downarrow} = \uparrow$. We determine the parameters $\Delta_{j\alpha}$ 
self-consistently  
by the gap equations
\begin{equation}\label{eq: GapVectorDefiningEquation}
    \begin{split}
        &\Delta_{j\sigma} = U\langle n_{j\sigma}\rangle,\\
        &\Delta_{j+} = \Delta_{j-}^* = -U\langle c_{j\uparrow}^\dagger c_{j\downarrow} \rangle.    
    \end{split}
\end{equation}
The decoupling in Eq.~\eqref{eq: GapVectorDefiningEquation} captures both the Hartree- ($\Delta_{j,\sigma}$) and Fock ($\Delta_{j,\pm}$) 
contributions and allows us to describe 
arbitrary spin and charge ordering patterns.  
The local charge- and spin expectation values are given by
\begin{equation}
 \begin{split}
  & \langle n_j\rangle = \frac{\Delta_{j\uparrow}+\Delta_{j\downarrow}}{U}, \\
  & \langle S^z_j\rangle=\frac{\Delta_{j\uparrow}-\Delta_{j\downarrow}}{2U} , \\
  & \langle S^x_j\rangle = -\frac{\Delta_{j+}+\Delta_{j-}}{2U} , \\
  & \langle S^y_j\rangle = -\frac{\Delta_{j+}-\Delta_{j-}}{2i U} ,
 \end{split}
\end{equation}
with $n_j = n_{j\uparrow} + n_{j\downarrow}$ and $\Vec{S}_j = \frac{1}{2} c^\dagger_j  \Vec{\sigma} c_j$, where $\vec{\sigma}$ is the vector of 
Pauli matrices.

We solve the mean-field equations on a finite lattice consisting of $N$ unit cells  
and periodic boundary conditions. 
The lattice is set up with parallel boundaries. The unit cells of single- and bilayer graphene contain two and four atoms, respectively. Thus, in total, the 
number of sites 
in our calculation is $\mathcal{N} = 2N$ for single-layer and $\mathcal{N} = 4N$  
for bilayer graphene.
We typically perform the real space calculations on lattices with $18\times 18$ unit cells for single-layer and $12 \times 12$ unit cells for bilayer graphene, 
i.e., we have $648$ and $576$ sites, respectively. 

With these specifications the mean-field  Hamiltonian  
can be written as
\begin{equation} \label{MeanFieldHamburger}
 H^\mathrm{MF} = H_0 + H_\mathrm{int}^\mathrm{MF} =
 \sum_{j,j'} \sum_{\sigma,\sigma'}
 c^\dagger_{j\sigma}\mathcal{H}_{jj'}^{\sigma\sigma'}c_{j'\sigma'} + \mbox{const},
\end{equation}
where $\mathcal{H}_{jj'}^{\sigma\sigma'}\in \mathbb{C}^{2\mathcal{N}\times2\mathcal{N}}$ is a square matrix with $4\mathcal{N}$ real parameters to be determined self-consistently ($\Delta_{j\uparrow}$, $\Delta_{j\downarrow}$, $\mathrm{Re}\Delta_{j+}$, and $\mathrm{Im}\Delta_{j+}$ for each lattice site $j$).
The expectation values of eq. \eqref{eq: GapVectorDefiningEquation} 
are then given by
\begin{equation} \label{eq: exp value T!=0}
    \langle c^\dagger_{j\sigma}c_{j\sigma'}\rangle = \sum_{\ell=1}^{2\mathcal{N}} (v^\ell_{j\sigma})^* v^\ell_{j\sigma'} f(\eps_\ell-\mu),
\end{equation}
where $f(x)=1/(1+e^{x/T})$ is the Fermi function, $T$ the temperature, $\eps_\ell$ are the eigenvalues of the matrix $\mathcal{H}_{jj'}^{\sigma\sigma'}$ from Eq.~\eqref{MeanFieldHamburger}, and $v^\ell_{j\sigma}$ is the normalized eigenvector corresponding to the $\ell$-th eigenvalue. 
This formulation allows us to perform the mean-field approximation also at finite temperatures.

Furthermore, we keep the total average of the electron density $n$ fixed in our calculation by enforcing the constraint 
\begin{equation} \label{eq: Find chemical potential}
    n=\frac{1}{\mathcal{N}}\sum_{j\sigma}\langle n_{j\sigma}\rangle=\frac{1}{\mathcal{N}}\sum_{\ell=1}^{2\mathcal{N}} f(\eps_\ell-\mu),
\end{equation}
To achieve this, we adjust the chemical potential $\mu$ in each iteration of the self-consistency loop.

Finally, when we obtain a convergent solution of the self-consistent mean-field equations, we calculate the free energy and determine the lowest-energy state via %
\begin{equation}\label{eq: free energy}
 \begin{split}
 F/\mathcal{N} = & -\frac{T}{\mathcal{N}}
 \sum_\ell \log(1 + e^{-\beta (\epsilon_\ell - \mu)}) \\
 & -\frac{1}{\mathcal{N}U}\sum_j
 \left( \Delta_{j\uparrow}\Delta_{j\downarrow} - \Delta_{j-}\Delta_{j+}\right) +
 \mu n.
 \end{split}
\end{equation}
We perform this Hartree-Fock procedure for varying temperature, density, and displacement field. To ensure that we do not converge to a local minimum, we repeat each calculation  
several times from random initial spin and charge patterns and after that from the (converged) states of adjacent points in the phase diagram. 

With a few exceptions, we 
consider 
calculations on a 
fixed lattice size. This means that the exact phase boundaries are subject to finite size effects. The influence of these kinds of finite size effects was analyzed in greater detail for the Hubbard model on a square lattice in \cite{Scholle2023,ScholleSpiralToStripe}. For an even more precise picture of the phase diagram, one could repeat the calculations for various lattice sizes comparing their free energies or constrain the spin orders in order to work directly in the thermodynamic limit. For our purposes, we  
intend to give an 
overview of the regimes where different phases are dominant without calculating sharp phase boundaries, 
and leave their exact determination for future study. We still perform checks on 18x18 and 24x24 lattices for several points for BBG to ensure a correct convergence. 

\subsection{Random Phase Approximation}
We additionally performed calculations using a random phase approximation (RPA) 
in the thermodynamic limit to compare the mean-field critical temperature $T^*$ for the onset of magnetic order. 
For this we 
introduce the Green's function of the non-interacting system  
and calculate the static spin susceptibility 
\begin{equation}\label{eq: RPA susceptibility}
\begin{split}
 \chi(\bq) &= \chi_0(\bq)
 \big[ \mathbb{1} - \Gamma_0 \chi_0(\bq) \big]^{-1} \, , \\
 \chi_0^{\alpha\beta}(\bq) &= \int_k G_0^{\alpha \beta}(k + q)
G_0^{\beta \alpha}(k),\\
G_0(k) &= \left( i\nu +\mu n -H_{0,\bk}  \right)^{-1},
 \end{split}
\end{equation}
where we introduced the shorthand notation $\int_k=T\sum_{i\nu} \int_{\bk \in \text{BZ}} d\bk \sqrt{3}/(8 \pi^2)$ and $k = (i\nu,\bk)$, $q = (0,\bq)$ contain a fermionic Matsubara frequency $\nu$ and a wavevector $\bk$ or $\bq$. 

We assume that right below the critical temperature $T^*$, the spin order will be given by a single contributing mode with wave vector $\bQ$ \cite{ScholleSpiralToStripe}, corresponding to a diverging $\chi(\bQ)$.
To determine $T^*$ and $\bQ$, we focus on $\bq$ along the high symmetry lines (i.e. $\Gamma - M - K -\Gamma$), and calculate the maximal critical temperature $T_\bq^*$ 
where at least one eigenvalue of  ${\big[\mathbb{1}_4 - \Gamma_0 \chi_0(\bq)\big]}$ becomes 0 so that 
the denominator of Eq.~\eqref{eq: RPA susceptibility} vanishes. We then determine the mean-field critical temperature as
\begin{equation}
    T^* = \max_\bq T_\bq^*,
\end{equation}
and $\bQ$ as the corresponding $\bq$, where $T_\bq^*$ becomes maximal. Note that the chemical potential needs to be calculated for each set of filling and temperature.

\subsection{Classification of real-space states}

Before we discuss the phase diagrams,  
we give an overview of the types of magnetic states we find and how we categorize them. We perform the classification of states  
based on the pattern of the spin expectation values. 
This classification is inspired by similar classifications for spin patterns on the square lattice \cite{Sachdev2019,Scholle2023}. In agreement with \cite{Sachdev2019,ZacharKivelson1998}, we generally find that the charge pattern of our states follows the spin pattern 
according to
\begin{equation}\label{eq: ChargeModulation}
\langle \rho_j \rangle \propto
\left\langle \vec{S}_j \right \rangle ^2 + \text{const.}
\end{equation}
Note that in general this relation only holds for small spin expectation values, since it was derived from a low-order Landau theory \cite{ZacharKivelson1998}.

\begin{figure}
    \centering
    \includegraphics[width=0.48\textwidth]{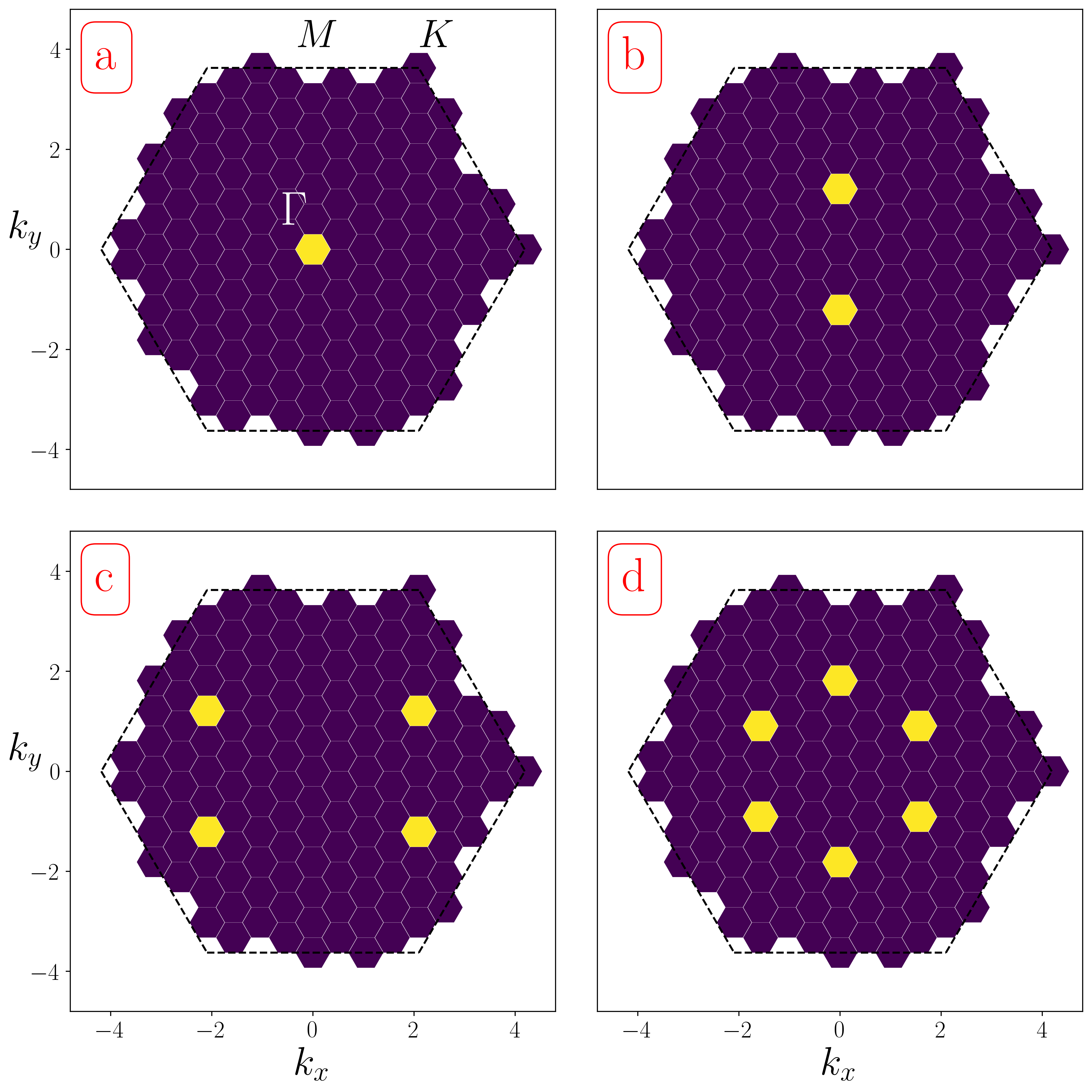}
    \caption{The four most common Fourier-patterns we find in our real-space Hartree-Fock calculations. 
    a) The spin order has only a single peak at $\bQ = 0$, indicating a resulting N\'eel or ferromagnetic order depending on the sublattice structure. 
    b) 
    There is a single contributing Fourier mode at $\bq = \pm \bQ$, corresponding to incommensurate single mode stripe order. 
    c) The spin order consists of two symmetry-related Fourier modes, which we mainly find in the two orthogonal stripe ordered phase. We here show the special case where they have the same amplitude. 
    d) 
    Three symmetry-related 
    Fourier-modes occur with the same amplitude. 
    For example, an order with three orthogonal  or collinear stripes 
    shows this Fourier pattern. As a special case, the three Fourier modes can occur at the three $M$ points.}
    \label{fig: FourierWeights}
\end{figure}

To classify
a magnetic state, we consider the Fourier-transformation of the real-space spin pattern
\begin{equation}
    \left\langle \vec{S}_{j} \right \rangle =\left\langle \vec{S}_{m,\alpha} \right \rangle = \sum_\bq \vec{S}_{\bq,\alpha} e^{i \bq \boldsymbol{r}_m},
\end{equation}
where we
specify a site $j$ by the  
the unit cell ($m$) at position $\boldsymbol{r}_m$ and the  
sublattice ($\alpha$).
Since the real-space spin pattern is real, 
the Fourier components of the spin expectation values satisfy $\vec{S}_{\bq,\alpha} = \vec{S}_{-\bq,\alpha}^*$. We find that the classification is the same for each of the sublattices, so we drop the extra index $\alpha$. 
The sum runs over the $N$ allowed momenta $\bq$ in the Brillouin zone given by the real-space lattice size. 
The Brillouin zone is discretized based on the $12 \times 12$ lattice we used to calculate our phase diagrams, 
i.e., there are $144$ allowed momenta for the spin pattern. In particular, this includes 
wave vectors for the spin orders 
at the $M$- or $K$-points.
In Fig. \ref{fig: FourierWeights}, we show the Fourier weight of the four most 
common types of spin orders we find in our numerical calculations. Generally, we obtain non-zero Fourier components at one, two, or three different $\bq$ and the related $-\bq$. They can be at high-symmetry points $\Gamma,K$, and $M$ or along high-symmetry lines $\Gamma-K$, $\Gamma-M$ or $K-M$. 
In addition, we determine the relative orientation of the Fourier-spins, 
since we can not completely classify a state with its  Fourier weights only.

\begin{figure*}
    \centering
    \includegraphics[width=1.\textwidth]{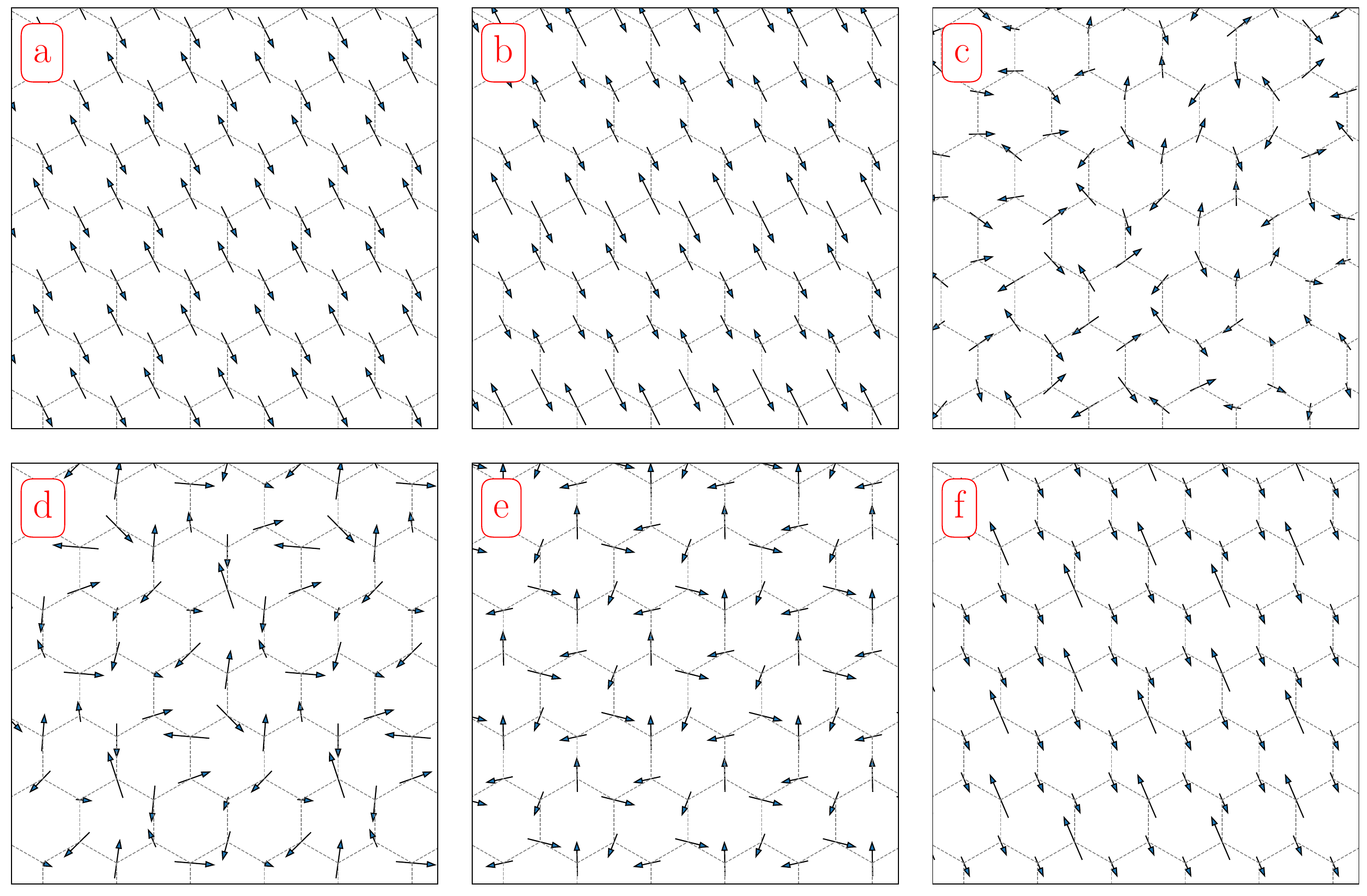}
    \caption{Example representatives of the real-space patterns of the most common spin orders 
    we find. We here show only one layer, but we found that in the case of BBG both layers display the same type of spin pattern for a given parameter set possibly with a relative phase.
    a) 
    N\'eel antiferromagnetism, 
    b)  
    single mode stripe order, 
    c)  
    two orthogonal stripes with the same amplitude,  
    d) 
    an order with three modes with same Fourier amplitude, which are neither orthogonal nor parallel stripes  
    e)  
    three orthogonal stripes 
    in the special case of commensurate wave vector  
    $\bQ_i = M_i$,   
    f)  
    three collinear stripes, again in the commensurate case $\bQ_i = M_i$. Orders in d) and e) are neither collinear nor coplanar and we only depict their projection on a plane. 
    }
    \label{fig: 6SpinOrders}
\end{figure*}

This leads us to the following categories for the most common spin orders we find. 
Note that the actual spin pattern we  
get in our Hartree-Fock calculations can be related to the 
states of the categories by a global SU(2) rotation or translations of the spin order.
We present 
example patterns 
in Fig. \ref{fig: 6SpinOrders}, illustrated by showing the real-space configuration of one graphene layer projected onto the $x-y-$plane in spin-space.

\subsubsection{Paramagnetism}
In the paramagnetic state, the average spin magnetization on every lattice site is zero,
$\langle \vec{S}_j \rangle = 0$
and the average electron density is uniform, $\langle n_j \rangle = n$.
\subsubsection{N\'eel or ferromagnetic order}
The spin order in each sublattice is given by 
\begin{equation}
    \vec{S}_{\boldsymbol{0}} = 
\mathcal{S}\begin{pmatrix}
 1 \\
 0 \\
 0
 \end{pmatrix}, \qquad 
 \vec{S}_\bq =  \vec{0} \quad \text{for} \quad \bq \neq \boldsymbol{0},
\end{equation}
with spin amplitude $S$.
This definition  
entails both ferromagnetism as well as  
N\'eel antiferromagnetism, depending on whether the spins on the different sublattices are arranged in a parallel or antiparallel way. Only if the spins are antiparallel, we classify the state as N\'eel order. In bilayer graphene, we additionally find that spins on the same site in different layers are antiparallel.  All spins on each sublattice have the same amplitude and thus we do not find charge modulations. The Fourier weight is shown in 
Fig. \ref{fig: FourierWeights} a) The real-space spin pattern is shown in Fig. \ref{fig: 6SpinOrders} a).
\subsubsection{Single mode stripe order}
We classify a state as single mode stripe order, if it can be described by only one mode $\bQ$ in momentum space, so that
\begin{equation}
    \vec{S}_\bQ = \vec{S}_{-\bQ}^* = \mathcal{S}\begin{pmatrix}
 1 \\
 0 \\
 0
 \end{pmatrix}, \qquad 
 \vec{S}_\bq =  \vec{0} \quad \text{for} \quad \bq \neq \pm\bQ.
\end{equation}
All the spins are pointing collinearly along one axis with their amplitude modulated with wavevector $\bQ$. The charge modulation on a sublattice will be to lowest order  proportional to $\langle n_m \rangle -n \propto \cos(2\bQ \cdot \mathbf{r}_m)$. The Fourier weight of single mode stripe order corresponds to Fig. \ref{fig: FourierWeights} b). Its real-space representation is shown in Fig. \ref{fig: 6SpinOrders} b). %
\subsubsection{Two orthogonal stripes}

If the spin order is given by 

\begin{equation}\label{eq: 2orthStripes}
\begin{split}
    \vec{S}_{\bQ_1} &= \vec{S}_{-\bQ_1}^* = \mathcal{S}_1\begin{pmatrix}
 1 \\
 0 \\
 0
 \end{pmatrix}, \\
    \vec{S}_{\bQ_2} &= \vec{S}_{-\bQ_2}^* = \mathcal{S}_2\begin{pmatrix}
 0 \\
 1 \\
 0
 \end{pmatrix},
\end{split}
\end{equation}

we classify a state as two orthogonal stripes. The wavevectors $\bQ_1$ and $\bQ_2$ have to be symmetry-related by a 
$C_3$ rotation. We often find a special case of this state where $\mathcal{S}_1 = \mathcal{S}_2$. We show the Fourier amplitudes of this state in Fig. \ref{fig: FourierWeights} c), where we present the common case %
with $\mathcal{S}_1 = \mathcal{S}_2$. In this case, the charge modulation follows $\langle n_m \rangle -n \propto \cos(2\bQ_1 \cdot \mathbf{r}_m) + \cos(2\bQ_2 \cdot \mathbf{r}_m)$.
Its real-space representation is shown in Fig. \ref{fig: 6SpinOrders} c).

\subsubsection{Three modes with same amplitude}
We classify a state as three modes with same amplitude, if
$S_\bq$ is nonzero only for three symmetry-related modes $\pm\bQ_1,\pm\bQ_2,\pm\bQ_3$, where $\bQ_1,\bQ_2$ and $\bQ_3$ are again related to each other by a $120^{\circ}$ rotation and 
\begin{equation}\label{eq: 3modessameamplitudes}
    |\vec{S}_{\bQ_1}| = |\vec{S}_{\bQ_2}| =|\vec{S}_{\bQ_3}|\,.
\end{equation}
We show an example in  Fig. \ref{fig: FourierWeights} d). We find several different types of spin orders in the form \eqref{eq: 3modessameamplitudes}, which are not necessarily coplanar. We show the $x-y-$projection of one example in Fig. \ref{fig: 6SpinOrders} d). 
We distinguish general states with three modes of the same amplitude from two special cases where the three spin modes are orthogonal or parallel.
\subsubsection{Orthogonal stripes}
Orthogonal stripes are a special case of three modes with the same amplitude, where the three spin amplitudes at $\bQ_1,\bQ_2,\bQ_3$ are orthogonal to each other, i.e.
\begin{equation}
\begin{split}
\vec{S}_{\bQ_1} &= \vec{S}^*_{-\bQ_1} = \mathcal{S}\begin{pmatrix}
 1 \\
 0 \\
 0
 \end{pmatrix} \\
 \vec{S}_{\bQ_2} &= \vec{S}^*_{-\bQ_2} = \mathcal{S}\begin{pmatrix}
 0 \\
 1 \\
 0
 \end{pmatrix}\\
 \vec{S}_{\bQ_3} &= \vec{S}^*_{-\bQ_3} = \mathcal{S}\begin{pmatrix}
 0 \\
 0 \\
 1
 \end{pmatrix}.
\end{split}
\end{equation}
The charge modulation is therefore proportional to $\langle n_m \rangle -n \propto \cos(2\bQ_1 \cdot \mathbf{r}_m) + \cos(2\bQ_2 \cdot \mathbf{r}_m) +\cos(2\bQ_3 \cdot \mathbf{r}_m)$.
Its Fourier-weight looks like that of the three same mode order, i.e.  Fig. \ref{fig: FourierWeights} d). 
Since this order is not coplanar, we show the projection of the spins to the $x-y-$plane in  Fig. \ref{fig: 6SpinOrders} e). In that panel, we  show the special case where $\bQ_i = M_i$. This state was previously discussed as a symmetry-broken state of the doped triangular or honeycomb lattice \cite{PhysRevLett.101.156402,PhysRevLett.105.266405,Li_2012,Nandkishore2012}. It forms a four-sublattice state where the spins on each sublattice point into the four corners of a tetrahedron.
\subsubsection{Collinear stripes}
Collinear stripes are a special case of three modes with the same amplitude, where the three spin amplitudes at $\bQ_1,\bQ_2,\bQ_3$ are collinear, i.e.
\begin{equation}
\vec{S}_{\bQ_1} = \vec{S}_{\bQ_2}
 =\vec{S}_{\bQ_3} =\mathcal{S}\begin{pmatrix}
 1 \\
 0 \\
 0
 \end{pmatrix},
\end{equation}
and its conjugate for $\bQ_i \leftrightarrow -\bQ_i$. The charge modulation is proportional to $\langle n_m \rangle -n \propto (\cos(\bQ_1 \cdot \mathbf{r}_m) + \cos(\bQ_2 \cdot \mathbf{r}_m) +\cos(\bQ_3 \cdot \mathbf{r}_m))^2$. In the special case $\bQ = M$, this state is equivalent to the collinearly ordered state proposed in \cite{Nandkishore2012}. Its Fourier weight is thus again represented by Fig. \ref{fig: FourierWeights} d). We show the spin order pattern in Fig. \ref{fig: 6SpinOrders} f).
\subsubsection{Other orders}
Any spin order we find which does not fit into one of the aforementioned categories, we classify as "other" spin orders.

Since we are performing our calculations on a finite system and completely unrestricted, there is the risk to converge to a local energy minimum, which is not the true lowest energy state of the system. Generally, the "other" orders tend to disappear when we increase the system size.  We therefore believe that the "other" orders are largely artifacts of convergences to local minima and finite size effects, similar to \cite{Scholle2023}.  

\section{Results}
\label{sec:results}

\subsection{Single Layer Graphene}
\label{sec:results1G}

For later reference, we 
perform Hartree-Fock and RPA calculations also for single layer graphene. 
We vary the temperature between $0$ and $0.5t$ in steps of $0.01$ and the density between 0.5 and 1.5 in steps of $\frac{1}{108}$. 
We choose an interaction strength of $U = 3t$ because,
on the one hand, it is big enough to find magnetism along a wide region of dopings and temperature, while on the other hand, it is still small enough for Hartree-Fock theory to provide meaningful results.

\begin{figure*}
    \centering
    \includegraphics[width=0.9\textwidth]{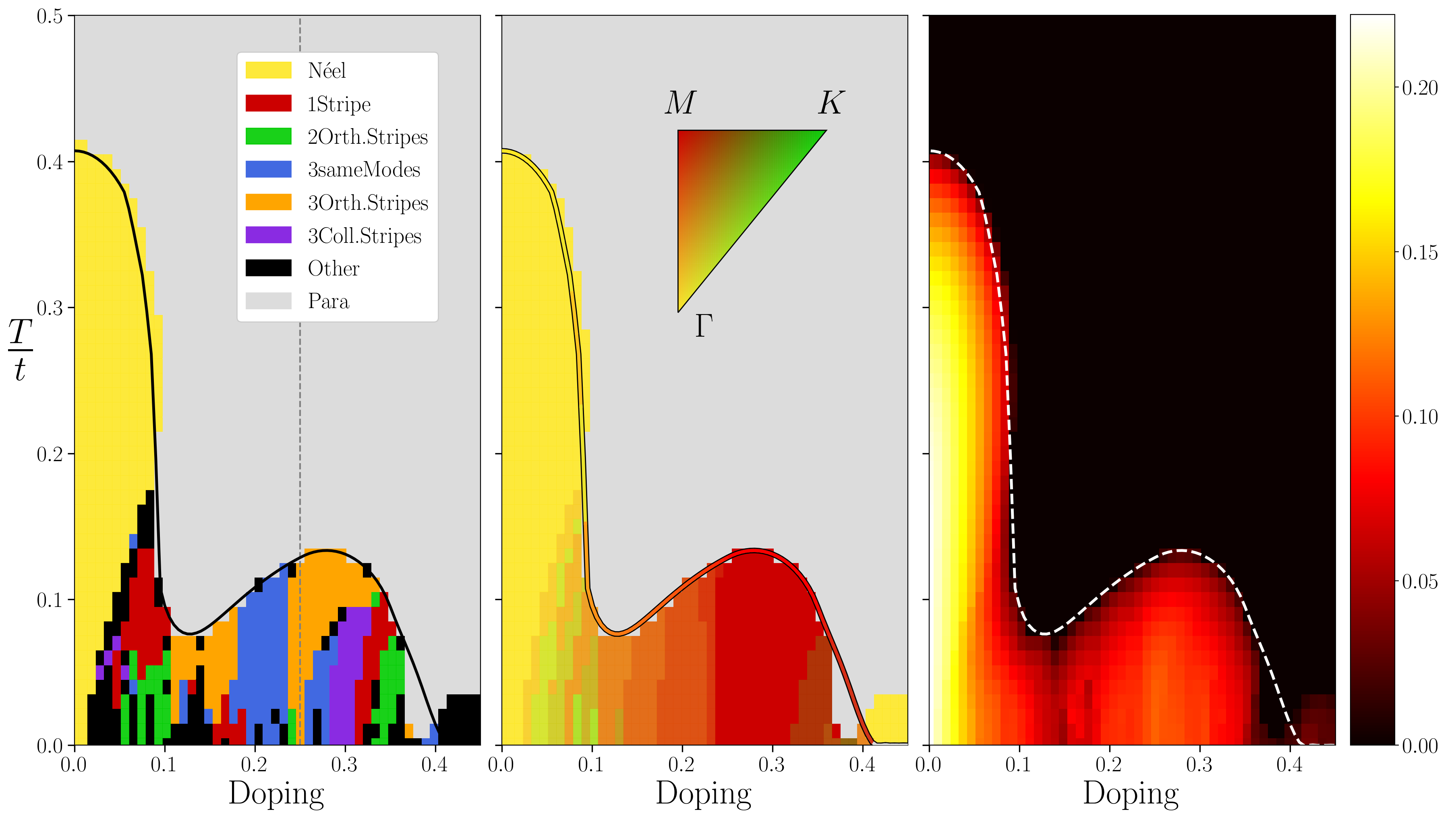}
    \caption{The Hartree-Fock phase diagram, main wavevector $\bQ$ (color scale in middle inset) and average magnetization $m$ (right color scale) of single layer graphene for $U = 3t$ as a function of temperature and doping. Calculations were done on $18 \times 18$ lattices. The grey dashed line at 25\% hole doping in the left panel indicates the Van Hove filling of the non-interacting model. The $T^*$ line and its corresponding $\bQ$ (color in the middle double line) were calculated in the thermodynamic limit using RPA. 
    }
    \label{fig: SingleLayerEverything}
\end{figure*}

Since single layer graphene is particle-hole symmetric, we only
consider hole doping. 
We present 
the phase diagram as function of doping $\delta$ and temperature, as well as the relevant spin ordering wavevectors $\bQ$  and the magnetization $m$ in Fig.~\ref{fig: SingleLayerEverything}. 
In general, we find striking similarities between BBG and single layer graphene, cf. Sec.~\ref{sec:resultsBBG}. 

The critical temperature as calculated in RPA 
coincides well with the critical temperature that we find in our real-space calculations, indicating that our system sizes are already big enough to give a good approximation of the thermodynamic limit. 
Around half-filling, we 
observe a N\'eel ordered dome consistent with previous studies, despite the vanishing density of states because the considered interaction $U=3t$ is above the critical interaction strength needed in mean-field theory \cite{PhysRevLett.100.146404,PhysRevB.73.214418,PhysRevB.82.115124,PhysRevB.103.205135,Meng2010,PhysRevB.101.125103,PhysRevB.81.115416,S.Sorella_1992}. The critical temperature $T^*$ is at 
a relative maximum because of particle-hole symmetry between conduction and valence bands $\epsilon_k^+=-\epsilon_k^-$ at charge neutrality. 
The regime of Ne\'el order is followed by an intermediate regime of single mode stripe order. 
A secondary dome arises around Van Hove filling because the logarithmic singularity of the density of states together with nesting at wave vector $M$, i.e., $\epsilon_{k+M}=-\epsilon_k$ boosts the critical temperature $T^*$. 
Within the secondary dome, we  
find an interplay of different types of three same modes, including orthogonal and collinear stripes, at dopings between $\delta = 0.15$ and $ \delta = 0.3$. Consistent with previous studies we find orthogonal stripes, i.e., a chiral spin density wave, at Van Hove doping \cite{PhysRevLett.101.156402,PhysRevLett.105.266405,Li_2012,Nandkishore2012}. Their competition with collinear stripes is influenced by temperature \cite{Nandkishore2012}, and more strongly by doping. 
For higher doping of  
0.3 to 0.35, we find a single mode stripe ordered regime
and for even higher dopings we see a small region of two orthogonal stripe order. 

In the middle panel of Fig.~\ref{fig: SingleLayerEverything}, we present the $\bQ$ dependence of the spin patterns as a function of doping and temperature. We color code the $\bQ$ dependence according to 
their location in the Brillouin zone (cf. inset). 
For the real-space Hartree-Fock calculations we use 
the $\bQ$-vector with the biggest weight of the corresponding Fourier-mode of the spin pattern.
The color of the outline is given by the $\bQ$ vector of the divergence 
in the RPA 
calculation. 
The $\bQ$ dependence of the magnetic orders 
is largely determined by the evolution of the Fermi surface with doping. 
For Fermi points and small pockets close to half-filling,  
we see  N\'eel order 
with $\bQ = 0$. For small dopings with Fermi pockets around $K,K'$, we find intermediate 
wave vectors with $\bQ$ moving along the $\Gamma-K$ line or $\Gamma - M$ line. 
The wave-vector fluctuations in this regime indicate that these states are close in energy.
In general, $|\bQ|$ increases for increasing doping or lower temperatures. In the secondary dome, the spin order reaches the $M$ point approximately at Van Hove doping, which corresponds to the nesting wave vector. Temperature effects slighlty shift the maximum away from Van Hove doping.  
At very high dopings, $\bQ$ shifts towards $K$ along the $M-K$ line. This evolution is in agreement with our RPA calculations of single layer graphene.

For each converged state, we calculated the average magnetization per site 
\begin{equation}
\label{eq:m}
    m = \frac{1}{\mathcal{N}} \sum_j \sqrt{\langle S^x_{j}\rangle^2+\langle S^y_{j}\rangle^2+\langle S^z_{j}\rangle^2},
\end{equation}
where the index $j$ runs over all sites of the system. We show our results in the right panel of Fig.~\ref{fig: SingleLayerEverything}.
The magnetization $m$ is highest at half-filling and for small $T$. The transition from the paramagnetic phase to the magnetically ordered phase is a second order transition within the resolution of our calculations. We also find a 
smooth magnetization profile for large ranges of doping. In particular, no magnetization jump occurs at the transition from N\'eel to incommensurate magnetic order, indicating 
a second order transition.
We see weak fluctuations in $m$ between the main and secondary magnetic dome. Since in the thermodynamic limit the magnetization should be stable up to the white RPA line, we believe these $m$ fluctuations are due to finite size effects of our lattice.

\subsection{Bernal Bilayer Graphene} 
\label{sec:resultsBBG}

In the following we state our results for BBG. 
In our calculations for BBG, we vary the temperature between $T = 0$ and $T = 0.5t$ in steps of $0.01t$ and the density between 0.5 and 1.5 in steps of $\frac{1}{144}$. 
We again choose an interaction strength of $U = 3t$. We are interested in the possible interaction-induced ordered states that mean-field theory provides to get a comprehensive overview as function of filling and temperature.

\subsubsection{Phase diagram}
\begin{figure}
    \centering
    \includegraphics[width=0.48\textwidth]{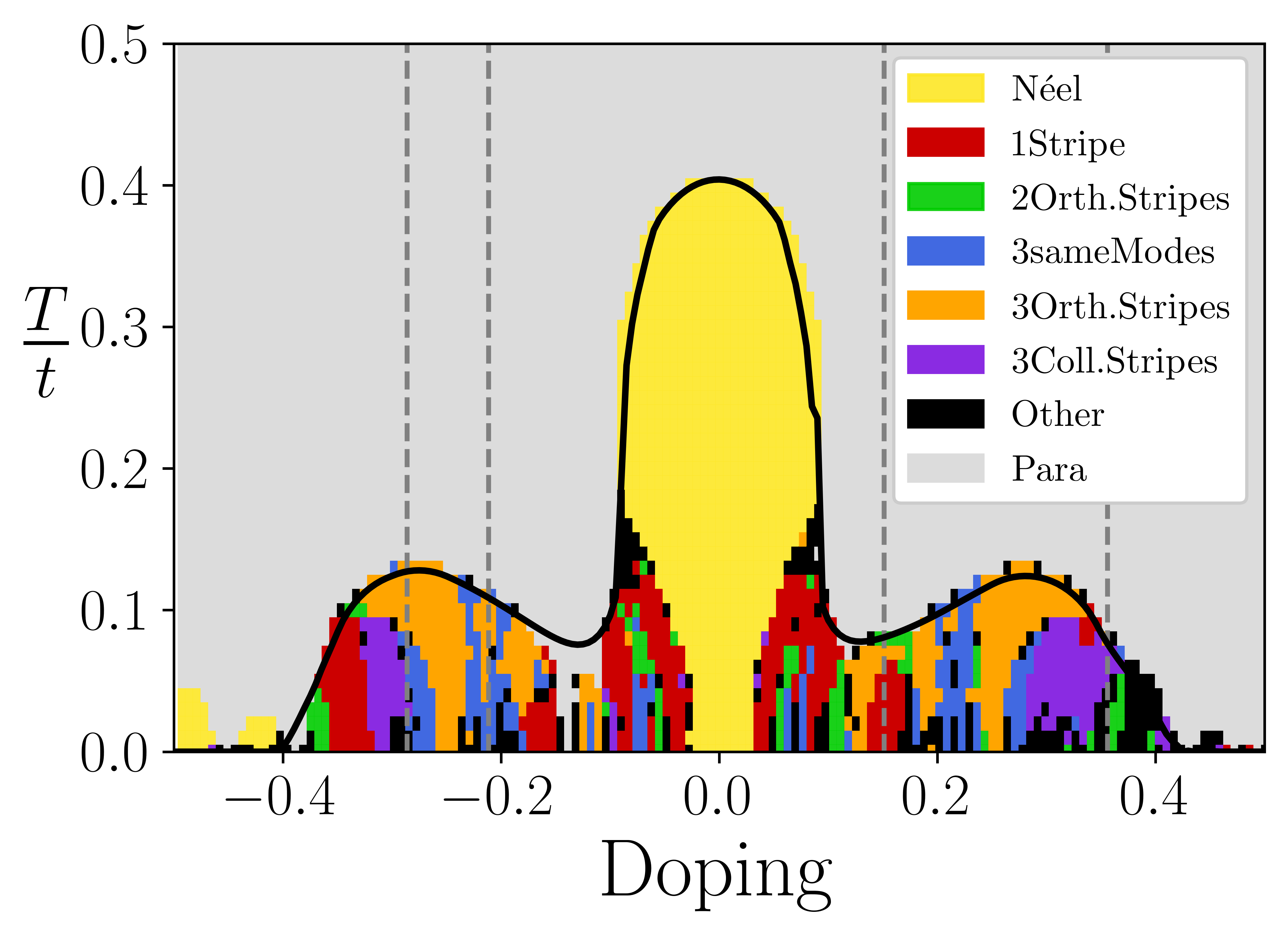}
    \caption{Mean-field phase diagram for the Hubbard model on Bernal bilayer graphene with $U = 3t$ as a function of 
    doping and temperature. Calculations were done  on $12 \times 12$ lattices. The black $T^*$ line was calculated in the thermodynamic limit using RPA. The grey dotted lines represent 
    Van Hove fillings of the non-interacting model at $n = 0.849, 0.644, 1.211$ and  $1.286$. 
    }
    \label{fig: PhaseDiagramBBGPhases}
\end{figure}

In Fig. \ref{fig: PhaseDiagramBBGPhases}, we show the phase diagram in hole/electron doping and temperature as obtained by our unrestricted real-space Hartree-Fock calculations. 
Note that in contrast to single-layer graphene the phase diagram is not perfectly particle-hole symmetric due to a finite $t_4$ hopping parameter. 
The RPA critical temperature 
again coincides well with the critical temperature that we find in the real-space calculations, 
i.e., the system sizes are 
a good approximation of the thermodynamic limit. 
The only exception is 
a small paramagnetic regime at doping $\delta\approx -0.1$ below the critical temperature. But we identify this as an artifact of the finite system size. When we repeat our calculations in this region on a $24\times 24$ lattice, we find a finite magnetic order parameter. 
Around half-filling, we see a magnetic dome with relatively high critical temperatures of up to $T^* \approx 0.40t$ in the phase diagram. As we increase the electron- or hole doping, the critical temperature first shrinks to $T^*\approx 0.08t$. Then, a secondary dome appears where the critical temperature rises again between 13 and 28 percent doping, peaking at $T^* \approx 0.13t$ and finally goes down to 0 at $\delta\approx \pm 0.4$. 
In highly doped BBG, there are two Van Hove energies from the two conduction (valence) bands. 
On the electron-doped side ($\delta<0$), the occurrence of the secondary dome 
coincides with a rise of the density of states and approximate nesting at Van Hove doping. On the hole-doped side, the maximum of $T^*$ is strongly shifted away from the individual Van Hove points. We suspect the reason of the strong shift to come from inter-band effects in the particle-hole susceptibility, besides temperature effects and the overlap of the two Van Hove peaks.

The dome around half filling predominantly corresponds to Néel order. 
At $T = 0$, the N\'eel order seems to extend over a small doping range, however, this is likely due to the finite size effects, as our system cannot resolve spin orders with very small incommensurabilities. 
For small dopings and low temperatures, a single mode stripe-ordered region emerges, extending 
from $\delta\approx 0.02$ to $\delta\approx 0.1$ and from $T = 0$ up to the critical temperature. 
For larger dopings $\delta\gtrsim 0.1$ in the region between the main magnetic dome and the secondary domes, the system undergoes several transitions between different stripe phases. As we describe below the wave vectors of magnetic orders remain stable when we increase the lattice size in our calculations. However, we attribute the absence of clear phase boundaries to finite size effects.

Within the secondary domes, we see an interplay of the "three-orthogonal-stripe", "three-same-modes", and "three-collinear-stripes" phases with wave vector $M$ similar to single-layer graphene. While this indicates that also in BBG the fermiology at Van Hove doping with approximate nesting is crucial for the order formation, the competition between the three phases is only resolved within a Hartree Fock calculation inside the symmetry-broken phases.  In particular, 
orthogonal stripes, i.e., chiral spin density waves, cover 
the largest 
area in the secondary domes.
Collinear stripes also cover extended regions for even 
larger hole- or electron dopings 
$\delta\approx\pm 0.3$. 
Generally, when phases extend over such big areas in the phase diagram, it is an indication that they will remain stable in the thermodynamic limit \cite{Scholle2023}. 
On the electron-doped side, we find an additional single mode stripe ordered regime for dopings even smaller than $-0.3$, which %
is not present on the hole-doped side. For very high dopings of $\delta\pm 0.4$, we see indications of a two-orthogonal-stripe order. In this area, the two orthogonal stripes have the same amplitude, i.e. $\mathcal{S}_1 = \mathcal{S}_2$ in Eq. \eqref{eq: 2orthStripes}.
Overall, the sequence of magnetic orders is very similar to the one of single-layer graphene, in particular on the electron doped side.

We also repeated the calculations on $18\times 18$ and $24 \times 24$ lattices along the $T = 0.04t$ and $T = 0.05t$ lines to estimate how stable the phase diagram will remain in the thermodynamic limit. %
In both of these lattice sizes, we find collinear stripes, orthogonal stripes and the phase with three same modes in the secondary domes, in qualitative agreement with Fig. \ref{fig: PhaseDiagramBBGPhases}. The collinear stripe phases appear stable. %
The phases of three orthogonal stripes and three same modes have slightly different boundaries for the bigger systems, indicating that their energies are very close and they are competing. 
We did not find a single mode stripe-ordered phase for dopings %
$\delta\leq-0.3$ on the electron-doped side for the bigger lattices. %
Instead the two orthogonal stripes take most of the phase space.
Based on these checks, we believe that the overall 
phase structure will remain stable even for bigger lattices.

\subsubsection{$\bQ$-dependence}

\begin{figure}
    \centering
    \includegraphics[width=0.48\textwidth]{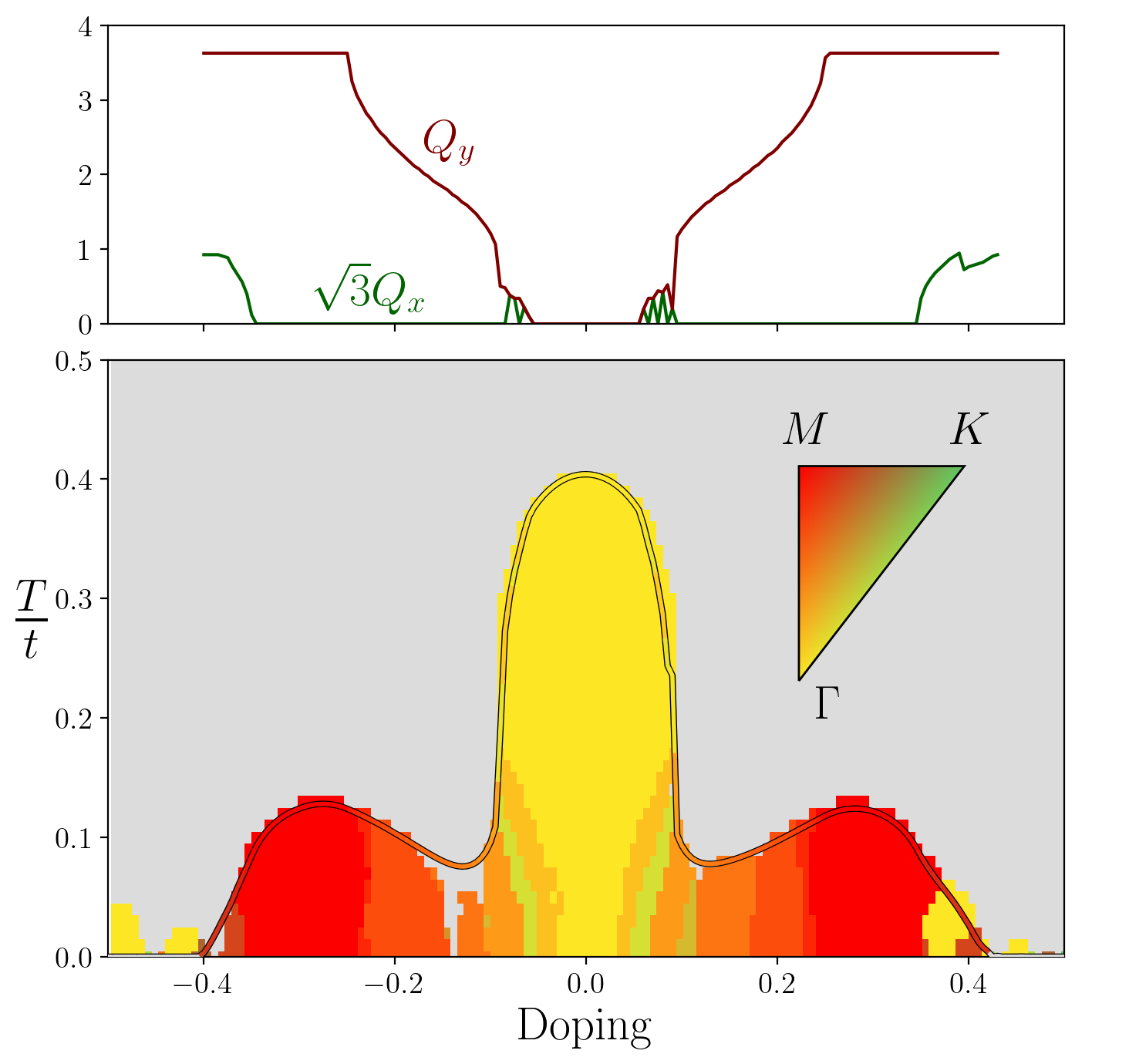}
    \caption{\textbf{Upper panel:} The $\bQ = (Q_x,Q_y)^T$, where we find a divergence of the RPA bubble in the thermodynamic limit. When the green line is zero, the divergence occurs at the $\Gamma - M$ line. When the green line equals the red line, the divergence occurs at the $\Gamma - K$ line. Otherwise it is at the $M-K$ line.
    \textbf{Lower panel}: The main $\bQ$ of the spin orders in real space. The color of the outline was calculated using RPA and corresponds to the upper panel.} 
    \label{fig: BBGQDependence}
\end{figure}

In 
Fig. \ref{fig: BBGQDependence}, we present the $\bQ$ dependence of the spin patterns corresponding to the phases in Fig.~\ref{fig: PhaseDiagramBBGPhases} as a function of doping and temperature.  
As before, we show the wave vector $\bQ$ of the divergence of the RPA bubble 
and the one of %
the real-space Hartree-Fock calculations 
given by the 
Fourier-mode of the spin pattern with the biggest weight. 
We see that within the resolution allowed by our $12\times 12$ lattice, at the critical temperature the $\bQ$ vectors of the RPA and the real-space Hartree-Fock calculations coincide, as expected.

Close to half-filling, we see 
a wave vector of $\bQ = \boldsymbol{0}$ 
for the N\'eel order.
Generally, we then see an increase of $|\bQ|$ as we increase electron or hole doping. 
Within our resolution, the transition from N\'eel order to the adjacent single mode stripe phase 
is consistent with a second-order transition as $\bQ$
increases by the smallest possible value allowed by the finite-size system. 
As function of temperature, we note an increase of  $|\bQ|$ as we lower the temperature in this region around half filling. 
We then have a short doping region ($\delta\approx~0.05$ - $\delta\approx~0.1$) 
where wave vectors change back and forth between $\Gamma-K$ and $\Gamma-M$ lines indicating that the  
energies of both magnetic orders are very close together.  
This reflects the changes between different phases we described above in this region of the phase diagram. 
While we cannot completely rule out finite-size effects, the $\bQ$ on the $\Gamma - K$ and $\Gamma - M$ lines  
remain stable 
in our checks on bigger lattices ($18 \times 18$ and $24 \times 24$).

For dopings 
$\delta\geq 0.1$, the wave vector settles along the $\Gamma-M$ line  
corresponding to an instability towards incommensurate 
spin density waves. At a doping of roughly $\delta\pm 0.25$ in the Van Hove region, $\bQ$ reaches the $M$ point and the magnetic order becomes commensurate. At even larger dopings 
$\delta\geq 0.35$, the ordering wave vector within RPA becomes again incommensurate close to $M$ along the $M-K$ line.
While this agrees with the Hartree-Fock results for electron doping, we find $\bQ = \boldsymbol{0}$ ferromagnetism for hole doping instead. 
However, when we repeat the calculation on bigger lattices, we 
obtain $\bQ=M$ or close to the $M-K$ line, in agreement with the predictions of the RPA. The reason why we find ferromagnetism in this region of the phase diagram is that a 12 $\times$ 12 lattice can not resolve the $M-K$ line sufficiently well and therefore $\bQ = 0$ ferromagnetic order is  
energetically the next best state. 
In the thermodynamic limit, we therefore do not expect ferromagnetic order to prevail. 
This is also why we have classified it as "other" order in the phase diagram.

\subsubsection{Magnetization}

\begin{figure}
    \centering
    \includegraphics[width=0.48\textwidth]{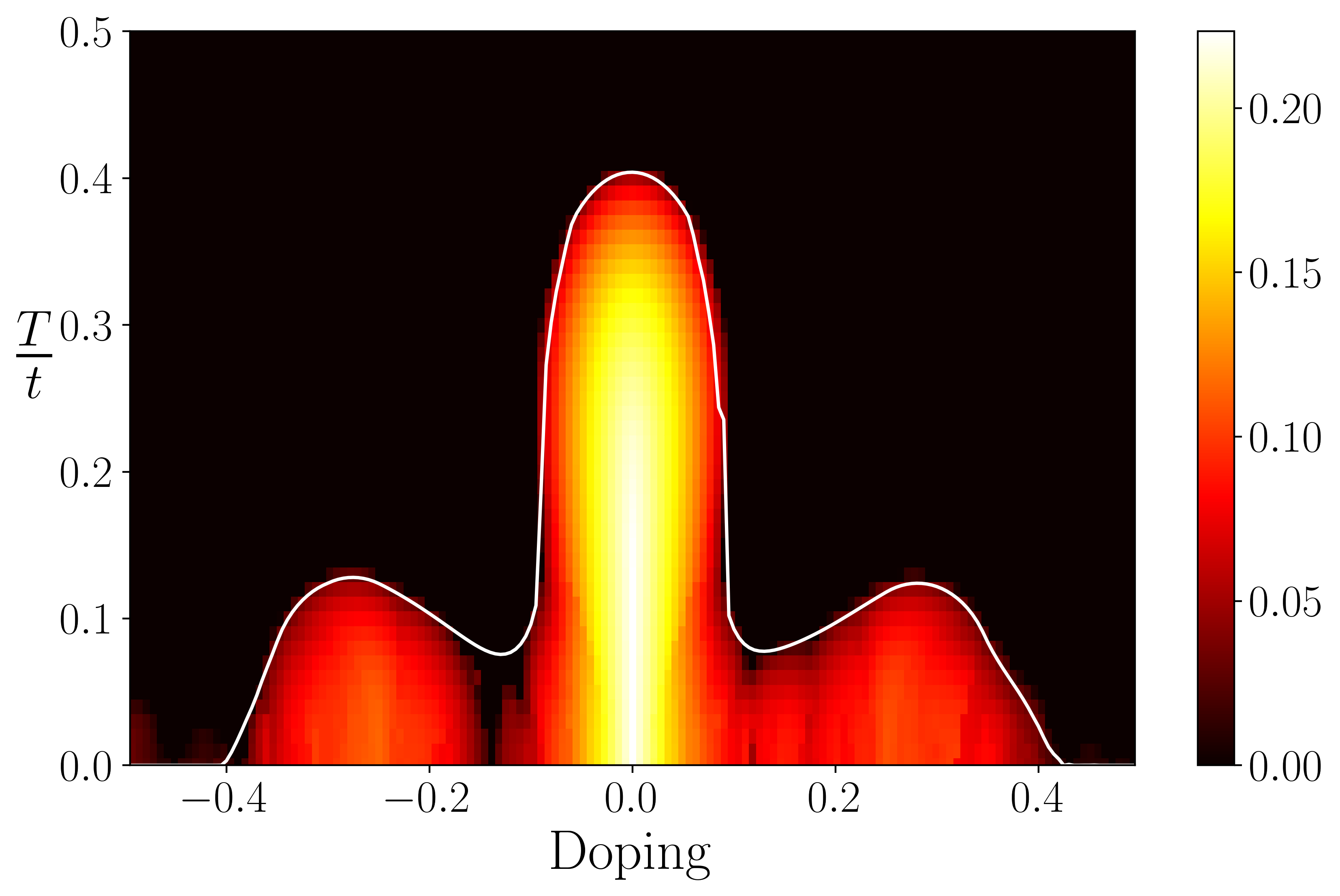}
    \caption{Average magnetization $m$ per site in Bernal bilayer graphene for $U = 3$ as a function of doping and temperature. The amplitudes are calculated using unrestricted Hartree-Fock and the white outline using RPA.}
    \label{fig: BBGMagnetization}
\end{figure}

For each converged state, we calculated the average magnetization per site according to Eq.~\eqref{eq:m}.
We show our results in Fig. \ref{fig: BBGMagnetization}. Similar to the single-layer case, the transition between the paramagnetic and the magnetic regime appears like a continuous second order transition within our density and temperature resolution. 
The magnetization takes its maximum at half-filling. In general, we do not observe jumps in the magnetization as we transition between different magnetic phases. 
This includes the transition from the N\'eel state to the adjacent single mode stripes. %
However, we do see fluctuations in $m$ in the following doping region between the main magnetic dome and the secondary magnetic domes, particularly on the electron-doped side. Since in the thermodynamic limit the magnetization should be stable up to the white RPA line, we believe these $m$ fluctuations are due to finite size effects. %

\subsubsection{Displacement field}

In addition, %
we also considered the effects of an externally applied displacement field, see Eq.~\ref{eq:D}. We note that we cannot resolve the cascade of transitions observed at smallest density regions as function of gate voltage. Instead, our motivation comes from intercalated BBG, which can lead to different charge densities in both layers. This can effectively be modeled by a term like Eq.~\ref{eq:D} in the Hamiltonian, which translates to different chemical potentials in both layers \cite{Yang_2022,PhDLink}. 
We perform calculations at $T = 0.01t$ to work at very low temperatures while having more numerical stability than for $T=0$. We choose a hole doping range from -0.1 to 0.5. We vary the displacement field from $D=0$ to $D = 2t$ in steps of $0.02t$ and set %
$U = 3t$.

\begin{figure}
    \centering
    \includegraphics[width=0.5\textwidth]{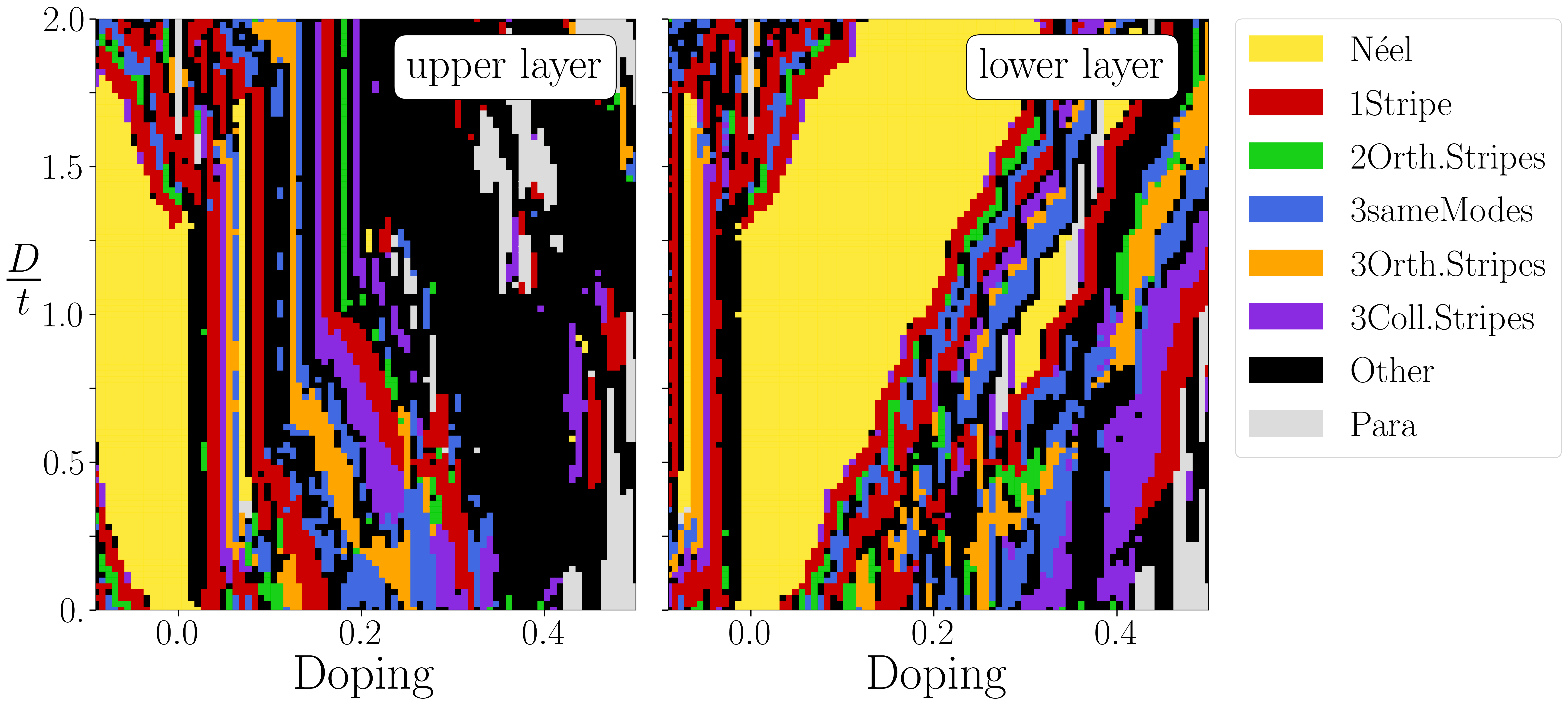}
    \caption{Phase diagram of BBG for an applied external magnetic field at $T = 0.01t$ and $U = 3$ as a function of doping and $D$. The left (right) panel shows the phase diagram for the sites on the upper (lower) layer. We can qualitatively describe the phase diagram by a cartoon picture of non-coupled graphene sheets, as shown in Fig. \ref{fig: Cartoon_DField}.}
    \label{fig: D_Diagram}
\end{figure}

In Fig. \ref{fig: D_Diagram} we show a phase diagram of BBG as a function of displacement field and doping. We see that now the two layers show qualitatively different behaviours. At half-filling, both layers exhibit N\'eel order up to displacement fields of $D \approx 1.3t$. We see that the lower layer has a wide region of N\'eel order in the hole doped regime. In this region, the upper layer instead shows parallel vertical lines of constant phase. These vertical lines of constant phases extend to the $D = 0$ line diagonally. In the lower layer, we also see these lines of constant phases in the electron doped regime. In the hole doped regime outside the N\'eel area, we again see diagonal lines of constant phase, this time with the diagonals going towards the top right. We note that the image is fairly pixelated, since we used less convergence time and still face the same finite size effects as we do for the phase diagrams in doping and temperature. Nevertheless, we believe that the phase diagram 
gives a qualitatively correct picture of the dominant field effects, which can be understood based on the phase diagram of single-layer graphene in Sec.~\ref{sec:results1G}.

\begin{figure}
    \centering
    \includegraphics[width=0.48\textwidth]{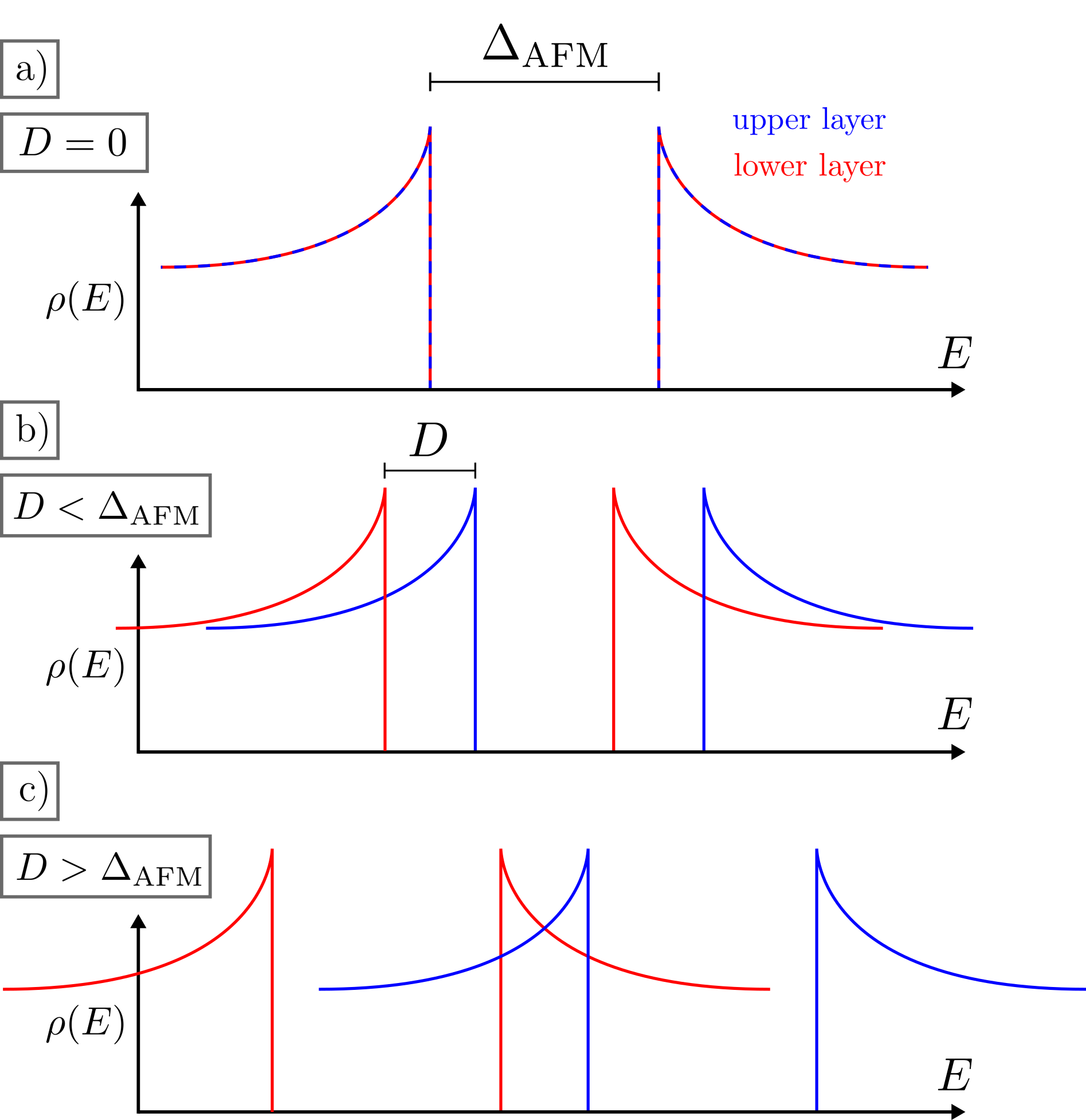}
    \caption{Cartoon picture of the density of states (DOS) $\rho$ of two non-connected layers of graphene. \textbf{Panel a)}: DOS of the system without displacement field. The DOS for the upper and the lower layer are degenerate. Both DOSs exhibit a gap $\Delta_{\text{AFM}}$ due to the N\'eel order. \textbf{Panel b)}: DOS after applying a small displacement field $D<\Delta_{\text{AFM}}$. The DOS of the upper and lower panel are shifted with respect to each other, but the magnetic gap still remains. \textbf{Panel c)}: DOS for $D>\Delta_{\text{AFM}}$. There is now no more magnetic gap remaining. }
    \label{fig: Cartoon_DField}
\end{figure}

To this end, we  
think of the system as two almost 
decoupled sheets of graphene which is a good approximation as long as $t\gg t_\perp$. 
The displacement field acts on the two layers like a chemical potential with flipped sign. 
A single layer of graphene in the N\'eel ordered state at half-filling and $U = 3t$ has a magnetic gap $\Delta_{\text{AFM}}$ in the density of states of $\Delta_{\text{AFM}}\approx 1.3t$, see Fig.~\ref{fig: Cartoon_DField} (a). At half-filling, we can see that both layers remain in the N\'eel state until $D \gtrapprox \Delta_{\text{AFM}}$. %
The reason is that, as long as $D<\Delta_{\text{AFM}}$, 
both layers remain half-filled (Fig.~\ref{fig: Cartoon_DField} (b)) and thus, exhibit perfect N\'eel order. 
When $D>\Delta_{\text{AFM}}$ exceeds the size of the magnetic gap, the different layers start to obtain different fillings. 
In this case the lower (upper) layer %
is filled slightly more (less) than half-filling, see Fig.~\ref{fig: Cartoon_DField} (c). Therefore, the order changes from a Ne\'el pattern to a single mode stripe phase following the phase diagram of  
BBG, Fig.~\ref{fig: PhaseDiagramBBGPhases}.

As we start hole doping the system the upper layer is depleted first, while the lower layer remains half-filled (for sufficiently big $D$ and $T = 0$), see
Fig.~\ref{fig: Cartoon_DField} (b). 
Effectively, the upper layer is then doped twice as much as the total system (e.g. when we consider 10\% hole doping  
the upper layer has 20\% and the lower layer 0\% hole doping). 
As a consequence, the lower layer remains Ne\'el ordered, while the upper layer 
shows incommensurate magnetic order.
The opposite effect happens as we electron dope the system. 
All additional electrons 
populate the lower layer first, while the upper layer remains at half filling. 
This mechanism is responsible for the extended N\'eel regions as well as the vertical lines of constant phases in the phase diagram of Fig. \ref{fig: D_Diagram}.

As we increase the doping even more, 
both layers become doped, however, with the lower (upper) layer remaining less doped across all hole (electron) dopings.  
This results in the diagonal lines of constant phases that can be seen in Fig. \ref{fig: D_Diagram}. They follow lines of constant doping of the individual layers. We can also identify a qualitatively similar sequence of phases. There is a fluctuating region of stripe phases with increasing doping, which transitions to larger regions of orders with three same modes, three collinear stripes and three orthogonal stripes.

\section{Conclusion}
\label{sec:conclude}

We performed unrestriced Hartree-Fock calculations on graphene and BBG lattice structures at moderate interaction strength $U = 3t$. The calculations were performed mainly on finite lattices with $18 \times 18$ unit cells for single layer graphene and $12 \times 12$ unit cells for BBG. Bigger lattice sizes were used for several selected points to double check our results and estimate finite size effects. Additionally, we performed RPA calculations directly in the thermodynamic limit for spin orders close to the critical temperature $T^*$. We classified the resulting states based on their magnetic pattern, calculated their average magnetization and showed the dominant Fourier weight of the spins.

We found that N\'eel order is  the dominant spin order close to half-filling for both single layer and Bernal bilayer graphene consistent with previous studies \cite{PhysRevLett.100.146404,PhysRevB.85.235408,PhysRevB.81.041401,PhysRevB.86.115447,PhysRevB.86.075467,PhysRevB.73.214418,PhysRevB.82.115124,PhysRevB.103.205135,Meng2010,PhysRevLett.109.126402,ZHANG20159,PhysRevB.101.125103,PhysRevB.81.115416,S.Sorella_1992}. Additionally, we studied the full phase diagram as the system gets doped away from half filling. We found that different stripe orders emerge with a wave vectors along $\Gamma-K$ and $\Gamma-M$ lines. In a second magnetic dome at $|\delta| \approx 0.3$, there is a competition of 3 orthogonal stripes, 3 collinear stripes and other phases with 3 Fourier modes with the same amplitude. 
These correspond to the previously discussed chiral (3 orthogonal stripes) and collinear (3 collinear stripes) spin density waves at Van Hove filling \cite{Li_2012,PhysRevLett.101.156402,Nandkishore2012,PhysRevLett.105.266405}. We showed that they extend away from Van Hove filling and mapped out the regimes in doping and temperature where they appear.

The qualitative shape of the phase diagram of single layer graphene and BBG is very similar, as well as the magnetic incommensurability $\bQ$ and average magnetizations. We found a smooth magnetization profile and a smooth variation of the incommensurability. A commensurate $\bQ = M$ emerges in an extended region in the secondary domes. 

We also studied the effect of an external displacement field $D$ on BBG. %
At the same time, the displacement field can also be understood as modelling unequal potentials in both layers, which is an effect that can occur in intercalation in BBG \cite{Yang_2022,PhDLink}. 
Motivated by this, we calculate a phase diagram %
for high doping levels as function of %
$D$,
which can qualitatively be explained by modeling BBG as two almost disconnected sheets of single layer graphene.

In future works, %
it will be interesting to extend the Hartree-Fock  analysis of this paper to clarify the %
exact nature of the phase boundaries, especially in the secondary dome. %
This can be done by limiting the allowed spin orders to the dominant spin orders that %
we found 
here and work directly in momentum space. %
This can also enable an investigation of the very small doping regime relevant for 
the recently observed cascade of phase transitions \cite{Li_2012,PhysRevLett.101.156402,Nandkishore2012,PhysRevLett.105.266405}. 
Finally, spin fluctuations can be taken into account by transforming the (long-range ordered) magnetic state into a fluctuating magnet with short-range (and potentially topological) order. This can be achieved by extending SU(2) gauge theories, incorporating Néel- and spiral-ordered chargons \cite{Scheurer2018, Sachdev2019review, Bonetti2022gauge}, to the new lattice geometry and also considering magnetic stripe order.
Experimentally, the different magnetic phases we find can be detected by scanning tunneling measurements, where a direct comparison with the Fourier modes and real space charge patterns, which follow the magnetisation of the local spin patterns, is possible. In this context, it will be interesting to further analyse the effect of extended Van Hove singularities observed in intercalated graphene samples \cite{PhysRevLett.104.136803,PhysRevB.100.121407,PhysRevLett.125.176403,PhysRevB.102.125141,HOVHSreview}. Our results demonstrate that intercalation of graphene and BBG can lead to interesting interaction-induced phases. 

\section{Acknowledgements}
We thank Walter Metzner, Andrey Chubukov and Pietro Bonetti for several interesting and fruitful discussions.

\newpage
\bibliography{main.bib}

\begin{thebibliography}{82}%
\makeatletter
\providecommand \@ifxundefined [1]{%
 \@ifx{#1\undefined}
}%
\providecommand \@ifnum [1]{%
 \ifnum #1\expandafter \@firstoftwo
 \else \expandafter \@secondoftwo
 \fi
}%
\providecommand \@ifx [1]{%
 \ifx #1\expandafter \@firstoftwo
 \else \expandafter \@secondoftwo
 \fi
}%
\providecommand \natexlab [1]{#1}%
\providecommand \enquote  [1]{``#1''}%
\providecommand \bibnamefont  [1]{#1}%
\providecommand \bibfnamefont [1]{#1}%
\providecommand \citenamefont [1]{#1}%
\providecommand \href@noop [0]{\@secondoftwo}%
\providecommand \href [0]{\begingroup \@sanitize@url \@href}%
\providecommand \@href[1]{\@@startlink{#1}\@@href}%
\providecommand \@@href[1]{\endgroup#1\@@endlink}%
\providecommand \@sanitize@url [0]{\catcode `\\12\catcode `\$12\catcode
  `\&12\catcode `\#12\catcode `\^12\catcode `\_12\catcode `\%12\relax}%
\providecommand \@@startlink[1]{}%
\providecommand \@@endlink[0]{}%
\providecommand \url  [0]{\begingroup\@sanitize@url \@url }%
\providecommand \@url [1]{\endgroup\@href {#1}{\urlprefix }}%
\providecommand \urlprefix  [0]{URL }%
\providecommand \Eprint [0]{\href }%
\providecommand \doibase [0]{https://doi.org/}%
\providecommand \selectlanguage [0]{\@gobble}%
\providecommand \bibinfo  [0]{\@secondoftwo}%
\providecommand \bibfield  [0]{\@secondoftwo}%
\providecommand \translation [1]{[#1]}%
\providecommand \BibitemOpen [0]{}%
\providecommand \bibitemStop [0]{}%
\providecommand \bibitemNoStop [0]{.\EOS\space}%
\providecommand \EOS [0]{\spacefactor3000\relax}%
\providecommand \BibitemShut  [1]{\csname bibitem#1\endcsname}%
\let\auto@bib@innerbib\@empty
\bibitem [{\citenamefont {Arovas}\ \emph {et~al.}(2022)\citenamefont {Arovas},
  \citenamefont {Berg}, \citenamefont {Kivelson},\ and\ \citenamefont
  {Raghu}}]{Arovas2022}%
  \BibitemOpen
  \bibfield  {author} {\bibinfo {author} {\bibfnamefont {D.~P.}\ \bibnamefont
  {Arovas}}, \bibinfo {author} {\bibfnamefont {E.}~\bibnamefont {Berg}},
  \bibinfo {author} {\bibfnamefont {S.~A.}\ \bibnamefont {Kivelson}},\ and\
  \bibinfo {author} {\bibfnamefont {S.}~\bibnamefont {Raghu}},\ }\bibfield
  {title} {\bibinfo {title} {{The Hubbard Model}},\ }\href
  {https://doi.org/10.1146/annurev-conmatphys-031620-102024} {\bibfield
  {journal} {\bibinfo  {journal} {Annu. Rev. Condens. Matter Phys.}\ }\textbf
  {\bibinfo {volume} {13}},\ \bibinfo {pages} {239} (\bibinfo {year}
  {2022})}\BibitemShut {NoStop}%
\bibitem [{\citenamefont {Qin}\ \emph {et~al.}(2022)\citenamefont {Qin},
  \citenamefont {Sch\"afer}, \citenamefont {Andergassen}, \citenamefont
  {Corboz},\ and\ \citenamefont {Gull}}]{Qin2022}%
  \BibitemOpen
  \bibfield  {author} {\bibinfo {author} {\bibfnamefont {M.}~\bibnamefont
  {Qin}}, \bibinfo {author} {\bibfnamefont {T.}~\bibnamefont {Sch\"afer}},
  \bibinfo {author} {\bibfnamefont {S.}~\bibnamefont {Andergassen}}, \bibinfo
  {author} {\bibfnamefont {P.}~\bibnamefont {Corboz}},\ and\ \bibinfo {author}
  {\bibfnamefont {E.}~\bibnamefont {Gull}},\ }\bibfield  {title} {\bibinfo
  {title} {{The Hubbard Model: A Computational Perspective}},\ }\href
  {https://doi.org/10.1146/annurev-conmatphys-090921-033948} {\bibfield
  {journal} {\bibinfo  {journal} {Annu. Rev. Condens. Matter Phys.}\ }\textbf
  {\bibinfo {volume} {13}},\ \bibinfo {pages} {275} (\bibinfo {year}
  {2022})}\BibitemShut {NoStop}%
\bibitem [{\citenamefont {Shraiman}\ and\ \citenamefont
  {Siggia}(1989)}]{Shraiman1989}%
  \BibitemOpen
  \bibfield  {author} {\bibinfo {author} {\bibfnamefont {B.~I.}\ \bibnamefont
  {Shraiman}}\ and\ \bibinfo {author} {\bibfnamefont {E.~D.}\ \bibnamefont
  {Siggia}},\ }\bibfield  {title} {\bibinfo {title} {{Spiral phase of a doped
  quantum antiferromagnet}},\ }\href
  {https://doi.org/10.1103/PhysRevLett.62.1564} {\bibfield  {journal} {\bibinfo
   {journal} {Phys. Rev. Lett.}\ }\textbf {\bibinfo {volume} {62}},\ \bibinfo
  {pages} {1564} (\bibinfo {year} {1989})}\BibitemShut {NoStop}%
\bibitem [{\citenamefont {Shraiman}\ and\ \citenamefont
  {Siggia}(1992)}]{Shraiman1992}%
  \BibitemOpen
  \bibfield  {author} {\bibinfo {author} {\bibfnamefont {B.~I.}\ \bibnamefont
  {Shraiman}}\ and\ \bibinfo {author} {\bibfnamefont {E.~D.}\ \bibnamefont
  {Siggia}},\ }\bibfield  {title} {\bibinfo {title} {{Excitation spectrum of
  the spiral state of a doped antiferromagnet}},\ }\href
  {https://doi.org/10.1103/PhysRevB.46.8305} {\bibfield  {journal} {\bibinfo
  {journal} {Phys. Rev. B}\ }\textbf {\bibinfo {volume} {46}},\ \bibinfo
  {pages} {8305} (\bibinfo {year} {1992})}\BibitemShut {NoStop}%
\bibitem [{\citenamefont {Chubukov}\ and\ \citenamefont
  {Frenkel}(1992)}]{Chubukov1992}%
  \BibitemOpen
  \bibfield  {author} {\bibinfo {author} {\bibfnamefont {A.~V.}\ \bibnamefont
  {Chubukov}}\ and\ \bibinfo {author} {\bibfnamefont {D.~M.}\ \bibnamefont
  {Frenkel}},\ }\bibfield  {title} {\bibinfo {title} {{Renormalized
  perturbation theory of magnetic instabilities in the two-dimensional Hubbard
  model at small doping}},\ }\href {https://doi.org/10.1103/PhysRevB.46.11884}
  {\bibfield  {journal} {\bibinfo  {journal} {Phys. Rev. B}\ }\textbf {\bibinfo
  {volume} {46}},\ \bibinfo {pages} {11884} (\bibinfo {year}
  {1992})}\BibitemShut {NoStop}%
\bibitem [{\citenamefont {Chubukov}\ and\ \citenamefont
  {Musaelian}(1995)}]{Chubukov1995}%
  \BibitemOpen
  \bibfield  {author} {\bibinfo {author} {\bibfnamefont {A.~V.}\ \bibnamefont
  {Chubukov}}\ and\ \bibinfo {author} {\bibfnamefont {K.~A.}\ \bibnamefont
  {Musaelian}},\ }\bibfield  {title} {\bibinfo {title} {{Magnetic phases of the
  two-dimensional Hubbard model at low doping}},\ }\href
  {https://doi.org/10.1103/PhysRevB.51.12605} {\bibfield  {journal} {\bibinfo
  {journal} {Phys. Rev. B}\ }\textbf {\bibinfo {volume} {51}},\ \bibinfo
  {pages} {12605} (\bibinfo {year} {1995})}\BibitemShut {NoStop}%
\bibitem [{\citenamefont {Dombre}(1990)}]{Dombre1990}%
  \BibitemOpen
  \bibfield  {author} {\bibinfo {author} {\bibfnamefont {T.}~\bibnamefont
  {Dombre}},\ }\bibfield  {title} {\bibinfo {title} {{Modulated spiral phases
  in doped quantum antiferromagnets}},\ }\href
  {https://doi.org/10.1051/jphys:01990005109084700} {\bibfield  {journal}
  {\bibinfo  {journal} {J. Phys. France}\ }\textbf {\bibinfo {volume} {51}},\
  \bibinfo {pages} {847} (\bibinfo {year} {1990})}\BibitemShut {NoStop}%
\bibitem [{\citenamefont {Fr{\'{e}}sard}\ \emph {et~al.}(1991)\citenamefont
  {Fr{\'{e}}sard}, \citenamefont {Dzierzawa},\ and\ \citenamefont
  {W\"olfle}}]{Fresard1991}%
  \BibitemOpen
  \bibfield  {author} {\bibinfo {author} {\bibfnamefont {R.}~\bibnamefont
  {Fr{\'{e}}sard}}, \bibinfo {author} {\bibfnamefont {M.}~\bibnamefont
  {Dzierzawa}},\ and\ \bibinfo {author} {\bibfnamefont {P.}~\bibnamefont
  {W\"olfle}},\ }\bibfield  {title} {\bibinfo {title} {{Slave-Boson Approach to
  Spiral Magnetic Order in the Hubbard Model}},\ }\href
  {https://doi.org/10.1209/0295-5075/15/3/016} {\bibfield  {journal} {\bibinfo
  {journal} {Europhys. Lett.}\ }\textbf {\bibinfo {volume} {15}},\ \bibinfo
  {pages} {325} (\bibinfo {year} {1991})}\BibitemShut {NoStop}%
\bibitem [{\citenamefont {Kotov}\ and\ \citenamefont
  {Sushkov}(2004)}]{Kotov2004}%
  \BibitemOpen
  \bibfield  {author} {\bibinfo {author} {\bibfnamefont {V.~N.}\ \bibnamefont
  {Kotov}}\ and\ \bibinfo {author} {\bibfnamefont {O.~P.}\ \bibnamefont
  {Sushkov}},\ }\bibfield  {title} {\bibinfo {title} {{Stability of the spiral
  phase in the two-dimensional extended $t\text{\ensuremath{-}}J$ model}},\
  }\href {https://doi.org/10.1103/PhysRevB.70.195105} {\bibfield  {journal}
  {\bibinfo  {journal} {Phys. Rev. B}\ }\textbf {\bibinfo {volume} {70}},\
  \bibinfo {pages} {195105} (\bibinfo {year} {2004})}\BibitemShut {NoStop}%
\bibitem [{\citenamefont {Igoshev}\ \emph {et~al.}(2010)\citenamefont
  {Igoshev}, \citenamefont {Timirgazin}, \citenamefont {Katanin}, \citenamefont
  {Arzhnikov},\ and\ \citenamefont {Irkhin}}]{Igoshev2010}%
  \BibitemOpen
  \bibfield  {author} {\bibinfo {author} {\bibfnamefont {P.~A.}\ \bibnamefont
  {Igoshev}}, \bibinfo {author} {\bibfnamefont {M.~A.}\ \bibnamefont
  {Timirgazin}}, \bibinfo {author} {\bibfnamefont {A.~A.}\ \bibnamefont
  {Katanin}}, \bibinfo {author} {\bibfnamefont {A.~K.}\ \bibnamefont
  {Arzhnikov}},\ and\ \bibinfo {author} {\bibfnamefont {V.~Y.}\ \bibnamefont
  {Irkhin}},\ }\bibfield  {title} {\bibinfo {title} {{Incommensurate magnetic
  order and phase separation in the two-dimensional Hubbard model with nearest-
  and next-nearest-neighbor hopping}},\ }\href
  {https://doi.org/10.1103/PhysRevB.81.094407} {\bibfield  {journal} {\bibinfo
  {journal} {Phys. Rev. B}\ }\textbf {\bibinfo {volume} {81}},\ \bibinfo
  {pages} {094407} (\bibinfo {year} {2010})}\BibitemShut {NoStop}%
\bibitem [{\citenamefont {Igoshev}\ \emph {et~al.}(2015)\citenamefont
  {Igoshev}, \citenamefont {Timirgazin}, \citenamefont {Gilmutdinov},
  \citenamefont {Arzhnikov},\ and\ \citenamefont {Irkhin}}]{Igoshev2015}%
  \BibitemOpen
  \bibfield  {author} {\bibinfo {author} {\bibfnamefont {P.~A.}\ \bibnamefont
  {Igoshev}}, \bibinfo {author} {\bibfnamefont {M.~A.}\ \bibnamefont
  {Timirgazin}}, \bibinfo {author} {\bibfnamefont {V.~F.}\ \bibnamefont
  {Gilmutdinov}}, \bibinfo {author} {\bibfnamefont {A.~K.}\ \bibnamefont
  {Arzhnikov}},\ and\ \bibinfo {author} {\bibfnamefont {V.~Y.}\ \bibnamefont
  {Irkhin}},\ }\bibfield  {title} {\bibinfo {title} {{Spiral magnetism in the
  single-band Hubbard model: the Hartree–Fock and slave-boson approaches}},\
  }\href {https://doi.org/10.1088/0953-8984/27/44/446002} {\bibfield  {journal}
  {\bibinfo  {journal} {Journal of Physics: Condensed Matter}\ }\textbf
  {\bibinfo {volume} {27}},\ \bibinfo {pages} {446002} (\bibinfo {year}
  {2015})}\BibitemShut {NoStop}%
\bibitem [{\citenamefont {Yamase}\ \emph {et~al.}(2016)\citenamefont {Yamase},
  \citenamefont {Eberlein},\ and\ \citenamefont {Metzner}}]{Yamase2016}%
  \BibitemOpen
  \bibfield  {author} {\bibinfo {author} {\bibfnamefont {H.}~\bibnamefont
  {Yamase}}, \bibinfo {author} {\bibfnamefont {A.}~\bibnamefont {Eberlein}},\
  and\ \bibinfo {author} {\bibfnamefont {W.}~\bibnamefont {Metzner}},\
  }\bibfield  {title} {\bibinfo {title} {{C}oexistence of {I}ncommensurate
  {M}agnetism and {S}uperconductivity in the two-dimensional {H}ubbard
  {M}odel},\ }\href {https://doi.org/10.1103/PhysRevLett.116.096402} {\bibfield
   {journal} {\bibinfo  {journal} {Phys. Rev. Lett.}\ }\textbf {\bibinfo
  {volume} {116}},\ \bibinfo {pages} {096402} (\bibinfo {year}
  {2016})}\BibitemShut {NoStop}%
\bibitem [{\citenamefont {Eberlein}\ \emph {et~al.}(2016)\citenamefont
  {Eberlein}, \citenamefont {Metzner}, \citenamefont {Sachdev},\ and\
  \citenamefont {Yamase}}]{Eberlein2016}%
  \BibitemOpen
  \bibfield  {author} {\bibinfo {author} {\bibfnamefont {A.}~\bibnamefont
  {Eberlein}}, \bibinfo {author} {\bibfnamefont {W.}~\bibnamefont {Metzner}},
  \bibinfo {author} {\bibfnamefont {S.}~\bibnamefont {Sachdev}},\ and\ \bibinfo
  {author} {\bibfnamefont {H.}~\bibnamefont {Yamase}},\ }\bibfield  {title}
  {\bibinfo {title} {Fermi surface reconstruction and drop in the hall number
  due to spiral antiferromagnetism in high-${T}_{c}$ cuprates},\ }\href
  {https://doi.org/10.1103/PhysRevLett.117.187001} {\bibfield  {journal}
  {\bibinfo  {journal} {Phys. Rev. Lett.}\ }\textbf {\bibinfo {volume} {117}},\
  \bibinfo {pages} {187001} (\bibinfo {year} {2016})}\BibitemShut {NoStop}%
\bibitem [{\citenamefont {Mitscherling}\ and\ \citenamefont
  {Metzner}(2018)}]{Mitscherling2018}%
  \BibitemOpen
  \bibfield  {author} {\bibinfo {author} {\bibfnamefont {J.}~\bibnamefont
  {Mitscherling}}\ and\ \bibinfo {author} {\bibfnamefont {W.}~\bibnamefont
  {Metzner}},\ }\bibfield  {title} {\bibinfo {title} {Longitudinal conductivity
  and hall coefficient in two-dimensional metals with spiral magnetic order},\
  }\href {https://doi.org/10.1103/PhysRevB.98.195126} {\bibfield  {journal}
  {\bibinfo  {journal} {Phys. Rev. B}\ }\textbf {\bibinfo {volume} {98}},\
  \bibinfo {pages} {195126} (\bibinfo {year} {2018})}\BibitemShut {NoStop}%
\bibitem [{\citenamefont {Bonetti}\ \emph {et~al.}(2020)\citenamefont
  {Bonetti}, \citenamefont {Mitscherling}, \citenamefont {Vilardi},\ and\
  \citenamefont {Metzner}}]{Bonetti2020a}%
  \BibitemOpen
  \bibfield  {author} {\bibinfo {author} {\bibfnamefont {P.~M.}\ \bibnamefont
  {Bonetti}}, \bibinfo {author} {\bibfnamefont {J.}~\bibnamefont
  {Mitscherling}}, \bibinfo {author} {\bibfnamefont {D.}~\bibnamefont
  {Vilardi}},\ and\ \bibinfo {author} {\bibfnamefont {W.}~\bibnamefont
  {Metzner}},\ }\bibfield  {title} {\bibinfo {title} {{Charge carrier drop at
  the onset of pseudogap behavior in the two-dimensional Hubbard model}},\
  }\href {https://doi.org/10.1103/PhysRevB.101.165142} {\bibfield  {journal}
  {\bibinfo  {journal} {Phys. Rev. B}\ }\textbf {\bibinfo {volume} {101}},\
  \bibinfo {pages} {165142} (\bibinfo {year} {2020})}\BibitemShut {NoStop}%
\bibitem [{\citenamefont {Schulz}(1989)}]{Schulz1989}%
  \BibitemOpen
  \bibfield  {author} {\bibinfo {author} {\bibfnamefont {H.~J.}\ \bibnamefont
  {Schulz}},\ }\bibfield  {title} {\bibinfo {title} {{Domain walls in a doped
  antiferromagnet}},\ }\href
  {https://doi.org/10.1051/jphys:0198900500180283300} {\bibfield  {journal}
  {\bibinfo  {journal} {J. Phys. France}\ }\textbf {\bibinfo {volume} {50}},\
  \bibinfo {pages} {2833} (\bibinfo {year} {1989})}\BibitemShut {NoStop}%
\bibitem [{\citenamefont {Zaanen}\ and\ \citenamefont
  {Gunnarsson}(1989)}]{Zaanen1989}%
  \BibitemOpen
  \bibfield  {author} {\bibinfo {author} {\bibfnamefont {J.}~\bibnamefont
  {Zaanen}}\ and\ \bibinfo {author} {\bibfnamefont {O.}~\bibnamefont
  {Gunnarsson}},\ }\bibfield  {title} {\bibinfo {title} {{Charged magnetic
  domain lines and the magnetism of high-${T}_{c}$ oxides}},\ }\href
  {https://doi.org/10.1103/PhysRevB.40.7391} {\bibfield  {journal} {\bibinfo
  {journal} {Phys. Rev. B}\ }\textbf {\bibinfo {volume} {40}},\ \bibinfo
  {pages} {7391} (\bibinfo {year} {1989})}\BibitemShut {NoStop}%
\bibitem [{\citenamefont {Machida}(1989)}]{Machida1989}%
  \BibitemOpen
  \bibfield  {author} {\bibinfo {author} {\bibfnamefont {K.}~\bibnamefont
  {Machida}},\ }\bibfield  {title} {\bibinfo {title} {{Magnetism in $\rm
  La_2CuO_4$ based compounds}},\ }\href
  {https://doi.org/https://doi.org/10.1016/0921-4534(89)90316-X} {\bibfield
  {journal} {\bibinfo  {journal} {Physica C: Superconductivity}\ }\textbf
  {\bibinfo {volume} {158}},\ \bibinfo {pages} {192} (\bibinfo {year}
  {1989})}\BibitemShut {NoStop}%
\bibitem [{\citenamefont {Poilblanc}\ and\ \citenamefont
  {Rice}(1989)}]{Poilblanc1989}%
  \BibitemOpen
  \bibfield  {author} {\bibinfo {author} {\bibfnamefont {D.}~\bibnamefont
  {Poilblanc}}\ and\ \bibinfo {author} {\bibfnamefont {T.~M.}\ \bibnamefont
  {Rice}},\ }\bibfield  {title} {\bibinfo {title} {{Charged solitons in the
  Hartree-Fock approximation to the large-U Hubbard model}},\ }\href
  {https://doi.org/10.1103/PhysRevB.39.9749} {\bibfield  {journal} {\bibinfo
  {journal} {Phys. Rev. B}\ }\textbf {\bibinfo {volume} {39}},\ \bibinfo
  {pages} {9749} (\bibinfo {year} {1989})}\BibitemShut {NoStop}%
\bibitem [{\citenamefont {Schulz}(1990)}]{Schulz1990}%
  \BibitemOpen
  \bibfield  {author} {\bibinfo {author} {\bibfnamefont {H.~J.}\ \bibnamefont
  {Schulz}},\ }\bibfield  {title} {\bibinfo {title} {{Incommensurate
  antiferromagnetism in the two-dimensional Hubbard model}},\ }\href
  {https://doi.org/10.1103/PhysRevLett.64.1445} {\bibfield  {journal} {\bibinfo
   {journal} {Phys. Rev. Lett.}\ }\textbf {\bibinfo {volume} {64}},\ \bibinfo
  {pages} {1445} (\bibinfo {year} {1990})}\BibitemShut {NoStop}%
\bibitem [{\citenamefont {Kato}\ \emph {et~al.}(1990)\citenamefont {Kato},
  \citenamefont {Machida}, \citenamefont {Nakanishi},\ and\ \citenamefont
  {Fujita}}]{Kato1990}%
  \BibitemOpen
  \bibfield  {author} {\bibinfo {author} {\bibfnamefont {M.}~\bibnamefont
  {Kato}}, \bibinfo {author} {\bibfnamefont {K.}~\bibnamefont {Machida}},
  \bibinfo {author} {\bibfnamefont {H.}~\bibnamefont {Nakanishi}},\ and\
  \bibinfo {author} {\bibfnamefont {M.}~\bibnamefont {Fujita}},\ }\bibfield
  {title} {\bibinfo {title} {{Soliton Lattice Modulation of Incommensurate Spin
  Density Wave in Two Dimensional Hubbard Model -A Mean Field Study}},\ }\href
  {https://doi.org/10.1143/JPSJ.59.1047} {\bibfield  {journal} {\bibinfo
  {journal} {J. Phys. Soc. Jpn.}\ }\textbf {\bibinfo {volume} {59}},\ \bibinfo
  {pages} {1047} (\bibinfo {year} {1990})}\BibitemShut {NoStop}%
\bibitem [{\citenamefont {Seibold}\ \emph {et~al.}(1998)\citenamefont
  {Seibold}, \citenamefont {Castellani}, \citenamefont {Di~Castro},\ and\
  \citenamefont {Grilli}}]{Seibold1998}%
  \BibitemOpen
  \bibfield  {author} {\bibinfo {author} {\bibfnamefont {G.}~\bibnamefont
  {Seibold}}, \bibinfo {author} {\bibfnamefont {C.}~\bibnamefont {Castellani}},
  \bibinfo {author} {\bibfnamefont {C.}~\bibnamefont {Di~Castro}},\ and\
  \bibinfo {author} {\bibfnamefont {M.}~\bibnamefont {Grilli}},\ }\bibfield
  {title} {\bibinfo {title} {{Striped phases in the two-dimensional Hubbard
  model with long-range Coulomb interaction}},\ }\href
  {https://doi.org/10.1103/PhysRevB.58.13506} {\bibfield  {journal} {\bibinfo
  {journal} {Phys. Rev. B}\ }\textbf {\bibinfo {volume} {58}},\ \bibinfo
  {pages} {13506} (\bibinfo {year} {1998})}\BibitemShut {NoStop}%
\bibitem [{\citenamefont {Fleck}\ \emph {et~al.}(2000)\citenamefont {Fleck},
  \citenamefont {Lichtenstein}, \citenamefont {Pavarini},\ and\ \citenamefont
  {Ole\ifmmode~\acute{s}\else \'{s}\fi{}}}]{Fleck2000}%
  \BibitemOpen
  \bibfield  {author} {\bibinfo {author} {\bibfnamefont {M.}~\bibnamefont
  {Fleck}}, \bibinfo {author} {\bibfnamefont {A.~I.}\ \bibnamefont
  {Lichtenstein}}, \bibinfo {author} {\bibfnamefont {E.}~\bibnamefont
  {Pavarini}},\ and\ \bibinfo {author} {\bibfnamefont {A.~M.}\ \bibnamefont
  {Ole\ifmmode~\acute{s}\else \'{s}\fi{}}},\ }\bibfield  {title} {\bibinfo
  {title} {{One-Dimensional Metallic Behavior of the Stripe Phase in
  ${\mathrm{La}}_{2\ensuremath{-}\mathit{x}}{\mathrm{Sr}}_{\mathit{x}}{\mathrm{CuO}}_{4}$}},\
  }\href {https://doi.org/10.1103/PhysRevLett.84.4962} {\bibfield  {journal}
  {\bibinfo  {journal} {Phys. Rev. Lett.}\ }\textbf {\bibinfo {volume} {84}},\
  \bibinfo {pages} {4962} (\bibinfo {year} {2000})}\BibitemShut {NoStop}%
\bibitem [{\citenamefont {Fleck}\ \emph {et~al.}(2001)\citenamefont {Fleck},
  \citenamefont {Lichtenstein},\ and\ \citenamefont {Ole\ifmmode~\acute{s}\else
  \'{s}\fi{}}}]{Fleck2001}%
  \BibitemOpen
  \bibfield  {author} {\bibinfo {author} {\bibfnamefont {M.}~\bibnamefont
  {Fleck}}, \bibinfo {author} {\bibfnamefont {A.~I.}\ \bibnamefont
  {Lichtenstein}},\ and\ \bibinfo {author} {\bibfnamefont {A.~M.}\ \bibnamefont
  {Ole\ifmmode~\acute{s}\else \'{s}\fi{}}},\ }\bibfield  {title} {\bibinfo
  {title} {{Spectral properties and pseudogap in the stripe phases of cuprate
  superconductors}},\ }\href {https://doi.org/10.1103/PhysRevB.64.134528}
  {\bibfield  {journal} {\bibinfo  {journal} {Phys. Rev. B}\ }\textbf {\bibinfo
  {volume} {64}},\ \bibinfo {pages} {134528} (\bibinfo {year}
  {2001})}\BibitemShut {NoStop}%
\bibitem [{\citenamefont {Raczkowski}\ and\ \citenamefont
  {Assaad}(2010)}]{Raczkowski2010}%
  \BibitemOpen
  \bibfield  {author} {\bibinfo {author} {\bibfnamefont {M.}~\bibnamefont
  {Raczkowski}}\ and\ \bibinfo {author} {\bibfnamefont {F.~F.}\ \bibnamefont
  {Assaad}},\ }\bibfield  {title} {\bibinfo {title} {{Melting of stripe phases
  and its signature in the single-particle spectral function}},\ }\href
  {https://doi.org/10.1103/PhysRevB.82.233101} {\bibfield  {journal} {\bibinfo
  {journal} {Phys. Rev. B}\ }\textbf {\bibinfo {volume} {82}},\ \bibinfo
  {pages} {233101} (\bibinfo {year} {2010})}\BibitemShut {NoStop}%
\bibitem [{\citenamefont {Timirgazin}\ \emph {et~al.}(2012)\citenamefont
  {Timirgazin}, \citenamefont {Arzhnikov},\ and\ \citenamefont
  {V.}}]{Timirgazin2012}%
  \BibitemOpen
  \bibfield  {author} {\bibinfo {author} {\bibfnamefont {M.~A.}\ \bibnamefont
  {Timirgazin}}, \bibinfo {author} {\bibfnamefont {M.~A.}\ \bibnamefont
  {Arzhnikov}},\ and\ \bibinfo {author} {\bibfnamefont {V.~A.}\ \bibnamefont
  {V.}},\ }\bibfield  {title} {\bibinfo {title} {{Incommensurate spin-density
  wave in two-dimensional Hubbard model}},\ }\href
  {https://doi.org/10.4028/www.scientific.net/SSP.190.67} {\bibfield  {journal}
  {\bibinfo  {journal} {Solid State Phenomena}\ }\textbf {\bibinfo {volume}
  {190}},\ \bibinfo {pages} {67} (\bibinfo {year} {2012})}\BibitemShut
  {NoStop}%
\bibitem [{\citenamefont {Peters}\ and\ \citenamefont
  {Kawakami}(2014)}]{Peters2014}%
  \BibitemOpen
  \bibfield  {author} {\bibinfo {author} {\bibfnamefont {R.}~\bibnamefont
  {Peters}}\ and\ \bibinfo {author} {\bibfnamefont {N.}~\bibnamefont
  {Kawakami}},\ }\bibfield  {title} {\bibinfo {title} {{Spin density waves in
  the Hubbard model: A DMFT approach}},\ }\href
  {https://doi.org/10.1103/PhysRevB.89.155134} {\bibfield  {journal} {\bibinfo
  {journal} {Phys. Rev. B}\ }\textbf {\bibinfo {volume} {89}},\ \bibinfo
  {pages} {155134} (\bibinfo {year} {2014})}\BibitemShut {NoStop}%
\bibitem [{\citenamefont {Matsuyama}\ and\ \citenamefont
  {Greensite}(2022)}]{Matsuyama2022}%
  \BibitemOpen
  \bibfield  {author} {\bibinfo {author} {\bibfnamefont {K.}~\bibnamefont
  {Matsuyama}}\ and\ \bibinfo {author} {\bibfnamefont {J.}~\bibnamefont
  {Greensite}},\ }\bibfield  {title} {\bibinfo {title} {{Multiplicity,
  localization, and domains in the Hartree–Fock ground state of the
  two-dimensional Hubbard model}},\ }\href
  {https://doi.org/https://doi.org/10.1016/j.aop.2022.168922} {\bibfield
  {journal} {\bibinfo  {journal} {Annals of Physics}\ }\textbf {\bibinfo
  {volume} {442}},\ \bibinfo {pages} {168922} (\bibinfo {year}
  {2022})}\BibitemShut {NoStop}%
\bibitem [{\citenamefont {Zheng}\ \emph {et~al.}(2017)\citenamefont {Zheng},
  \citenamefont {Chung}, \citenamefont {Corboz}, \citenamefont {Ehlers},
  \citenamefont {Qin}, \citenamefont {Noack}, \citenamefont {Shi},
  \citenamefont {White}, \citenamefont {Zhang},\ and\ \citenamefont
  {Chan}}]{Zheng2017}%
  \BibitemOpen
  \bibfield  {author} {\bibinfo {author} {\bibfnamefont {B.-X.}\ \bibnamefont
  {Zheng}}, \bibinfo {author} {\bibfnamefont {C.-M.}\ \bibnamefont {Chung}},
  \bibinfo {author} {\bibfnamefont {P.}~\bibnamefont {Corboz}}, \bibinfo
  {author} {\bibfnamefont {G.}~\bibnamefont {Ehlers}}, \bibinfo {author}
  {\bibfnamefont {M.-P.}\ \bibnamefont {Qin}}, \bibinfo {author} {\bibfnamefont
  {R.~M.}\ \bibnamefont {Noack}}, \bibinfo {author} {\bibfnamefont
  {H.}~\bibnamefont {Shi}}, \bibinfo {author} {\bibfnamefont {S.~R.}\
  \bibnamefont {White}}, \bibinfo {author} {\bibfnamefont {S.}~\bibnamefont
  {Zhang}},\ and\ \bibinfo {author} {\bibfnamefont {G.~K.-L.}\ \bibnamefont
  {Chan}},\ }\bibfield  {title} {\bibinfo {title} {{Stripe order in the
  underdoped region of the two-dimensional Hubbard model}},\ }\href
  {https://doi.org/10.1126/science.aam7127} {\bibfield  {journal} {\bibinfo
  {journal} {Science}\ }\textbf {\bibinfo {volume} {358}},\ \bibinfo {pages}
  {1155} (\bibinfo {year} {2017})}\BibitemShut {NoStop}%
\bibitem [{\citenamefont {Qin}\ \emph {et~al.}(2020)\citenamefont {Qin},
  \citenamefont {Chung}, \citenamefont {Shi}, \citenamefont {Vitali},
  \citenamefont {Hubig}, \citenamefont {Schollw\"ock}, \citenamefont {White},\
  and\ \citenamefont {Zhang}}]{Qin2020}%
  \BibitemOpen
  \bibfield  {author} {\bibinfo {author} {\bibfnamefont {M.}~\bibnamefont
  {Qin}}, \bibinfo {author} {\bibfnamefont {C.-M.}\ \bibnamefont {Chung}},
  \bibinfo {author} {\bibfnamefont {H.}~\bibnamefont {Shi}}, \bibinfo {author}
  {\bibfnamefont {E.}~\bibnamefont {Vitali}}, \bibinfo {author} {\bibfnamefont
  {C.}~\bibnamefont {Hubig}}, \bibinfo {author} {\bibfnamefont
  {U.}~\bibnamefont {Schollw\"ock}}, \bibinfo {author} {\bibfnamefont {S.~R.}\
  \bibnamefont {White}},\ and\ \bibinfo {author} {\bibfnamefont
  {S.}~\bibnamefont {Zhang}} (\bibinfo {collaboration} {Simons Collaboration on
  the Many-Electron Problem}),\ }\bibfield  {title} {\bibinfo {title} {{Absence
  of Superconductivity in the Pure Two-Dimensional Hubbard Model}},\ }\href
  {https://doi.org/10.1103/PhysRevX.10.031016} {\bibfield  {journal} {\bibinfo
  {journal} {Phys. Rev. X}\ }\textbf {\bibinfo {volume} {10}},\ \bibinfo
  {pages} {031016} (\bibinfo {year} {2020})}\BibitemShut {NoStop}%
\bibitem [{\citenamefont {Sch\"afer}\ \emph {et~al.}(2021)\citenamefont
  {Sch\"afer}, \citenamefont {Wentzell}, \citenamefont {\ifmmode~\check{S}\else
  \v{S}\fi{}imkovic}, \citenamefont {He}, \citenamefont {Hille}, \citenamefont
  {Klett}, \citenamefont {Eckhardt}, \citenamefont {Arzhang}, \citenamefont
  {Harkov}, \citenamefont {Le~R\'egent}, \citenamefont {Kirsch}, \citenamefont
  {Wang}, \citenamefont {Kim}, \citenamefont {Kozik}, \citenamefont {Stepanov},
  \citenamefont {Kauch}, \citenamefont {Andergassen}, \citenamefont {Hansmann},
  \citenamefont {Rohe}, \citenamefont {Vilk}, \citenamefont {LeBlanc},
  \citenamefont {Zhang}, \citenamefont {Tremblay}, \citenamefont {Ferrero},
  \citenamefont {Parcollet},\ and\ \citenamefont
  {Georges}}]{PhysRevX.11.011058}%
  \BibitemOpen
  \bibfield  {author} {\bibinfo {author} {\bibfnamefont {T.}~\bibnamefont
  {Sch\"afer}}, \bibinfo {author} {\bibfnamefont {N.}~\bibnamefont {Wentzell}},
  \bibinfo {author} {\bibfnamefont {F.}~\bibnamefont {\ifmmode~\check{S}\else
  \v{S}\fi{}imkovic}}, \bibinfo {author} {\bibfnamefont {Y.-Y.}\ \bibnamefont
  {He}}, \bibinfo {author} {\bibfnamefont {C.}~\bibnamefont {Hille}}, \bibinfo
  {author} {\bibfnamefont {M.}~\bibnamefont {Klett}}, \bibinfo {author}
  {\bibfnamefont {C.~J.}\ \bibnamefont {Eckhardt}}, \bibinfo {author}
  {\bibfnamefont {B.}~\bibnamefont {Arzhang}}, \bibinfo {author} {\bibfnamefont
  {V.}~\bibnamefont {Harkov}}, \bibinfo {author} {\bibfnamefont {F.~m. c.-M.}\
  \bibnamefont {Le~R\'egent}}, \bibinfo {author} {\bibfnamefont
  {A.}~\bibnamefont {Kirsch}}, \bibinfo {author} {\bibfnamefont
  {Y.}~\bibnamefont {Wang}}, \bibinfo {author} {\bibfnamefont {A.~J.}\
  \bibnamefont {Kim}}, \bibinfo {author} {\bibfnamefont {E.}~\bibnamefont
  {Kozik}}, \bibinfo {author} {\bibfnamefont {E.~A.}\ \bibnamefont {Stepanov}},
  \bibinfo {author} {\bibfnamefont {A.}~\bibnamefont {Kauch}}, \bibinfo
  {author} {\bibfnamefont {S.}~\bibnamefont {Andergassen}}, \bibinfo {author}
  {\bibfnamefont {P.}~\bibnamefont {Hansmann}}, \bibinfo {author}
  {\bibfnamefont {D.}~\bibnamefont {Rohe}}, \bibinfo {author} {\bibfnamefont
  {Y.~M.}\ \bibnamefont {Vilk}}, \bibinfo {author} {\bibfnamefont {J.~P.~F.}\
  \bibnamefont {LeBlanc}}, \bibinfo {author} {\bibfnamefont {S.}~\bibnamefont
  {Zhang}}, \bibinfo {author} {\bibfnamefont {A.-M.~S.}\ \bibnamefont
  {Tremblay}}, \bibinfo {author} {\bibfnamefont {M.}~\bibnamefont {Ferrero}},
  \bibinfo {author} {\bibfnamefont {O.}~\bibnamefont {Parcollet}},\ and\
  \bibinfo {author} {\bibfnamefont {A.}~\bibnamefont {Georges}},\ }\bibfield
  {title} {\bibinfo {title} {Tracking the footprints of spin fluctuations: A
  multimethod, multimessenger study of the two-dimensional hubbard model},\
  }\href {https://doi.org/10.1103/PhysRevX.11.011058} {\bibfield  {journal}
  {\bibinfo  {journal} {Phys. Rev. X}\ }\textbf {\bibinfo {volume} {11}},\
  \bibinfo {pages} {011058} (\bibinfo {year} {2021})}\BibitemShut {NoStop}%
\bibitem [{\citenamefont {Scholle}\ \emph {et~al.}(2023)\citenamefont
  {Scholle}, \citenamefont {Bonetti}, \citenamefont {Vilardi},\ and\
  \citenamefont {Metzner}}]{Scholle2023}%
  \BibitemOpen
  \bibfield  {author} {\bibinfo {author} {\bibfnamefont {R.}~\bibnamefont
  {Scholle}}, \bibinfo {author} {\bibfnamefont {P.~M.}\ \bibnamefont
  {Bonetti}}, \bibinfo {author} {\bibfnamefont {D.}~\bibnamefont {Vilardi}},\
  and\ \bibinfo {author} {\bibfnamefont {W.}~\bibnamefont {Metzner}},\
  }\bibfield  {title} {\bibinfo {title} {{Comprehensive mean-field analysis of
  magnetic and charge orders in the two-dimensional Hubbard model}},\ }\href
  {https://doi.org/10.1103/PhysRevB.108.035139} {\bibfield  {journal} {\bibinfo
   {journal} {Phys. Rev. B}\ }\textbf {\bibinfo {volume} {108}},\ \bibinfo
  {pages} {035139} (\bibinfo {year} {2023})}\BibitemShut {NoStop}%
\bibitem [{\citenamefont {Scholle}\ \emph {et~al.}(2024)\citenamefont
  {Scholle}, \citenamefont {Metzner}, \citenamefont {Vilardi},\ and\
  \citenamefont {Bonetti}}]{ScholleSpiralToStripe}%
  \BibitemOpen
  \bibfield  {author} {\bibinfo {author} {\bibfnamefont {R.}~\bibnamefont
  {Scholle}}, \bibinfo {author} {\bibfnamefont {W.}~\bibnamefont {Metzner}},
  \bibinfo {author} {\bibfnamefont {D.}~\bibnamefont {Vilardi}},\ and\ \bibinfo
  {author} {\bibfnamefont {P.~M.}\ \bibnamefont {Bonetti}},\ }\bibfield
  {title} {\bibinfo {title} {Spiral to stripe transition in the two-dimensional
  hubbard model},\ }\href {https://doi.org/10.1103/PhysRevB.109.235149}
  {\bibfield  {journal} {\bibinfo  {journal} {Phys. Rev. B}\ }\textbf {\bibinfo
  {volume} {109}},\ \bibinfo {pages} {235149} (\bibinfo {year}
  {2024})}\BibitemShut {NoStop}%
\bibitem [{\citenamefont {Scheurer}\ \emph {et~al.}(2018)\citenamefont
  {Scheurer}, \citenamefont {Chatterjee}, \citenamefont {Wu}, \citenamefont
  {Ferrero}, \citenamefont {Georges},\ and\ \citenamefont
  {Sachdev}}]{Scheurer2018}%
  \BibitemOpen
  \bibfield  {author} {\bibinfo {author} {\bibfnamefont {M.~S.}\ \bibnamefont
  {Scheurer}}, \bibinfo {author} {\bibfnamefont {S.}~\bibnamefont
  {Chatterjee}}, \bibinfo {author} {\bibfnamefont {W.}~\bibnamefont {Wu}},
  \bibinfo {author} {\bibfnamefont {M.}~\bibnamefont {Ferrero}}, \bibinfo
  {author} {\bibfnamefont {A.}~\bibnamefont {Georges}},\ and\ \bibinfo {author}
  {\bibfnamefont {S.}~\bibnamefont {Sachdev}},\ }\bibfield  {title} {\bibinfo
  {title} {{Topological order in the pseudogap metal}},\ }\href
  {https://doi.org/10.1073/pnas.1720580115} {\bibfield  {journal} {\bibinfo
  {journal} {Proc. Natl. Acad. Sci. USA}\ }\textbf {\bibinfo {volume} {115}},\
  \bibinfo {pages} {E3665} (\bibinfo {year} {2018})}\BibitemShut {NoStop}%
\bibitem [{\citenamefont {Sachdev}(2019)}]{Sachdev2019review}%
  \BibitemOpen
  \bibfield  {author} {\bibinfo {author} {\bibfnamefont {S.}~\bibnamefont
  {Sachdev}},\ }\bibfield  {title} {\bibinfo {title} {{Topological order,
  emergent gauge fields, and Fermi surface reconstruction}},\ }\href
  {https://doi.org/10.1088/1361-6633/aae110} {\bibfield  {journal} {\bibinfo
  {journal} {Rep. Prog. Phys.}\ }\textbf {\bibinfo {volume} {82}},\ \bibinfo
  {pages} {014001} (\bibinfo {year} {2019})}\BibitemShut {NoStop}%
\bibitem [{\citenamefont {Bonetti}\ and\ \citenamefont
  {Metzner}(2022)}]{Bonetti2022gauge}%
  \BibitemOpen
  \bibfield  {author} {\bibinfo {author} {\bibfnamefont {P.~M.}\ \bibnamefont
  {Bonetti}}\ and\ \bibinfo {author} {\bibfnamefont {W.}~\bibnamefont
  {Metzner}},\ }\bibfield  {title} {\bibinfo {title} {{SU(2) gauge theory of
  the pseudogap phase in the two-dimensional Hubbard model}},\ }\href
  {https://doi.org/10.1103/PhysRevB.106.205152} {\bibfield  {journal} {\bibinfo
   {journal} {Phys. Rev. B}\ }\textbf {\bibinfo {volume} {106}},\ \bibinfo
  {pages} {205152} (\bibinfo {year} {2022})}\BibitemShut {NoStop}%
\bibitem [{\citenamefont {Castro~Neto}\ \emph {et~al.}(2009)\citenamefont
  {Castro~Neto}, \citenamefont {Guinea}, \citenamefont {Peres}, \citenamefont
  {Novoselov},\ and\ \citenamefont {Geim}}]{RevModPhys.81.109}%
  \BibitemOpen
  \bibfield  {author} {\bibinfo {author} {\bibfnamefont {A.~H.}\ \bibnamefont
  {Castro~Neto}}, \bibinfo {author} {\bibfnamefont {F.}~\bibnamefont {Guinea}},
  \bibinfo {author} {\bibfnamefont {N.~M.~R.}\ \bibnamefont {Peres}}, \bibinfo
  {author} {\bibfnamefont {K.~S.}\ \bibnamefont {Novoselov}},\ and\ \bibinfo
  {author} {\bibfnamefont {A.~K.}\ \bibnamefont {Geim}},\ }\bibfield  {title}
  {\bibinfo {title} {The electronic properties of graphene},\ }\href
  {https://doi.org/10.1103/RevModPhys.81.109} {\bibfield  {journal} {\bibinfo
  {journal} {Rev. Mod. Phys.}\ }\textbf {\bibinfo {volume} {81}},\ \bibinfo
  {pages} {109} (\bibinfo {year} {2009})}\BibitemShut {NoStop}%
\bibitem [{\citenamefont {McCann}\ and\ \citenamefont
  {Koshino}(2013)}]{McCann_2013}%
  \BibitemOpen
  \bibfield  {author} {\bibinfo {author} {\bibfnamefont {E.}~\bibnamefont
  {McCann}}\ and\ \bibinfo {author} {\bibfnamefont {M.}~\bibnamefont
  {Koshino}},\ }\bibfield  {title} {\bibinfo {title} {The electronic properties
  of bilayer graphene},\ }\href {https://doi.org/10.1088/0034-4885/76/5/056503}
  {\bibfield  {journal} {\bibinfo  {journal} {Reports on Progress in Physics}\
  }\textbf {\bibinfo {volume} {76}},\ \bibinfo {pages} {056503} (\bibinfo
  {year} {2013})}\BibitemShut {NoStop}%
\bibitem [{\citenamefont {Honerkamp}(2008)}]{PhysRevLett.100.146404}%
  \BibitemOpen
  \bibfield  {author} {\bibinfo {author} {\bibfnamefont {C.}~\bibnamefont
  {Honerkamp}},\ }\bibfield  {title} {\bibinfo {title} {Density waves and
  cooper pairing on the honeycomb lattice},\ }\href
  {https://doi.org/10.1103/PhysRevLett.100.146404} {\bibfield  {journal}
  {\bibinfo  {journal} {Phys. Rev. Lett.}\ }\textbf {\bibinfo {volume} {100}},\
  \bibinfo {pages} {146404} (\bibinfo {year} {2008})}\BibitemShut {NoStop}%
\bibitem [{\citenamefont {Scherer}\ \emph {et~al.}(2012)\citenamefont
  {Scherer}, \citenamefont {Uebelacker},\ and\ \citenamefont
  {Honerkamp}}]{PhysRevB.85.235408}%
  \BibitemOpen
  \bibfield  {author} {\bibinfo {author} {\bibfnamefont {M.~M.}\ \bibnamefont
  {Scherer}}, \bibinfo {author} {\bibfnamefont {S.}~\bibnamefont
  {Uebelacker}},\ and\ \bibinfo {author} {\bibfnamefont {C.}~\bibnamefont
  {Honerkamp}},\ }\bibfield  {title} {\bibinfo {title} {Instabilities of
  interacting electrons on the honeycomb bilayer},\ }\href
  {https://doi.org/10.1103/PhysRevB.85.235408} {\bibfield  {journal} {\bibinfo
  {journal} {Phys. Rev. B}\ }\textbf {\bibinfo {volume} {85}},\ \bibinfo
  {pages} {235408} (\bibinfo {year} {2012})}\BibitemShut {NoStop}%
\bibitem [{\citenamefont {Vafek}\ and\ \citenamefont
  {Yang}(2010)}]{PhysRevB.81.041401}%
  \BibitemOpen
  \bibfield  {author} {\bibinfo {author} {\bibfnamefont {O.}~\bibnamefont
  {Vafek}}\ and\ \bibinfo {author} {\bibfnamefont {K.}~\bibnamefont {Yang}},\
  }\bibfield  {title} {\bibinfo {title} {Many-body instability of coulomb
  interacting bilayer graphene: Renormalization group approach},\ }\href
  {https://doi.org/10.1103/PhysRevB.81.041401} {\bibfield  {journal} {\bibinfo
  {journal} {Phys. Rev. B}\ }\textbf {\bibinfo {volume} {81}},\ \bibinfo
  {pages} {041401} (\bibinfo {year} {2010})}\BibitemShut {NoStop}%
\bibitem [{\citenamefont {Throckmorton}\ and\ \citenamefont
  {Vafek}(2012)}]{PhysRevB.86.115447}%
  \BibitemOpen
  \bibfield  {author} {\bibinfo {author} {\bibfnamefont {R.~E.}\ \bibnamefont
  {Throckmorton}}\ and\ \bibinfo {author} {\bibfnamefont {O.}~\bibnamefont
  {Vafek}},\ }\bibfield  {title} {\bibinfo {title} {Fermions on bilayer
  graphene: Symmetry breaking for $b=0$ and $\ensuremath{\nu}=0$},\ }\href
  {https://doi.org/10.1103/PhysRevB.86.115447} {\bibfield  {journal} {\bibinfo
  {journal} {Phys. Rev. B}\ }\textbf {\bibinfo {volume} {86}},\ \bibinfo
  {pages} {115447} (\bibinfo {year} {2012})}\BibitemShut {NoStop}%
\bibitem [{\citenamefont {Cvetkovic}\ \emph {et~al.}(2012)\citenamefont
  {Cvetkovic}, \citenamefont {Throckmorton},\ and\ \citenamefont
  {Vafek}}]{PhysRevB.86.075467}%
  \BibitemOpen
  \bibfield  {author} {\bibinfo {author} {\bibfnamefont {V.}~\bibnamefont
  {Cvetkovic}}, \bibinfo {author} {\bibfnamefont {R.~E.}\ \bibnamefont
  {Throckmorton}},\ and\ \bibinfo {author} {\bibfnamefont {O.}~\bibnamefont
  {Vafek}},\ }\bibfield  {title} {\bibinfo {title} {Electronic multicriticality
  in bilayer graphene},\ }\href {https://doi.org/10.1103/PhysRevB.86.075467}
  {\bibfield  {journal} {\bibinfo  {journal} {Phys. Rev. B}\ }\textbf {\bibinfo
  {volume} {86}},\ \bibinfo {pages} {075467} (\bibinfo {year}
  {2012})}\BibitemShut {NoStop}%
\bibitem [{\citenamefont {Nilsson}\ \emph {et~al.}(2006)\citenamefont
  {Nilsson}, \citenamefont {Castro~Neto}, \citenamefont {Peres},\ and\
  \citenamefont {Guinea}}]{PhysRevB.73.214418}%
  \BibitemOpen
  \bibfield  {author} {\bibinfo {author} {\bibfnamefont {J.}~\bibnamefont
  {Nilsson}}, \bibinfo {author} {\bibfnamefont {A.~H.}\ \bibnamefont
  {Castro~Neto}}, \bibinfo {author} {\bibfnamefont {N.~M.~R.}\ \bibnamefont
  {Peres}},\ and\ \bibinfo {author} {\bibfnamefont {F.}~\bibnamefont
  {Guinea}},\ }\bibfield  {title} {\bibinfo {title} {Electron-electron
  interactions and the phase diagram of a graphene bilayer},\ }\href
  {https://doi.org/10.1103/PhysRevB.73.214418} {\bibfield  {journal} {\bibinfo
  {journal} {Phys. Rev. B}\ }\textbf {\bibinfo {volume} {73}},\ \bibinfo
  {pages} {214418} (\bibinfo {year} {2006})}\BibitemShut {NoStop}%
\bibitem [{\citenamefont {Nandkishore}\ and\ \citenamefont
  {Levitov}(2010)}]{PhysRevB.82.115124}%
  \BibitemOpen
  \bibfield  {author} {\bibinfo {author} {\bibfnamefont {R.}~\bibnamefont
  {Nandkishore}}\ and\ \bibinfo {author} {\bibfnamefont {L.}~\bibnamefont
  {Levitov}},\ }\bibfield  {title} {\bibinfo {title} {Quantum anomalous hall
  state in bilayer graphene},\ }\href
  {https://doi.org/10.1103/PhysRevB.82.115124} {\bibfield  {journal} {\bibinfo
  {journal} {Phys. Rev. B}\ }\textbf {\bibinfo {volume} {82}},\ \bibinfo
  {pages} {115124} (\bibinfo {year} {2010})}\BibitemShut {NoStop}%
\bibitem [{\citenamefont {Szab\'o}\ and\ \citenamefont
  {Roy}(2021)}]{PhysRevB.103.205135}%
  \BibitemOpen
  \bibfield  {author} {\bibinfo {author} {\bibfnamefont {A.~L.}\ \bibnamefont
  {Szab\'o}}\ and\ \bibinfo {author} {\bibfnamefont {B.}~\bibnamefont {Roy}},\
  }\bibfield  {title} {\bibinfo {title} {Extended hubbard model in undoped and
  doped monolayer and bilayer graphene: Selection rules and organizing
  principle among competing orders},\ }\href
  {https://doi.org/10.1103/PhysRevB.103.205135} {\bibfield  {journal} {\bibinfo
   {journal} {Phys. Rev. B}\ }\textbf {\bibinfo {volume} {103}},\ \bibinfo
  {pages} {205135} (\bibinfo {year} {2021})}\BibitemShut {NoStop}%
\bibitem [{\citenamefont {Meng}\ \emph {et~al.}(2010)\citenamefont {Meng},
  \citenamefont {Lang}, \citenamefont {Wessel}, \citenamefont {Assaad},\ and\
  \citenamefont {Muramatsu}}]{Meng2010}%
  \BibitemOpen
  \bibfield  {author} {\bibinfo {author} {\bibfnamefont {Z.~Y.}\ \bibnamefont
  {Meng}}, \bibinfo {author} {\bibfnamefont {T.~C.}\ \bibnamefont {Lang}},
  \bibinfo {author} {\bibfnamefont {S.}~\bibnamefont {Wessel}}, \bibinfo
  {author} {\bibfnamefont {F.~F.}\ \bibnamefont {Assaad}},\ and\ \bibinfo
  {author} {\bibfnamefont {A.}~\bibnamefont {Muramatsu}},\ }\bibfield  {title}
  {\bibinfo {title} {Quantum spin liquid emerging in two-dimensional correlated
  dirac fermions},\ }\href {https://doi.org/10.1038/nature08942} {\bibfield
  {journal} {\bibinfo  {journal} {Nature}\ }\textbf {\bibinfo {volume} {464}},\
  \bibinfo {pages} {847} (\bibinfo {year} {2010})}\BibitemShut {NoStop}%
\bibitem [{\citenamefont {Lang}\ \emph {et~al.}(2012)\citenamefont {Lang},
  \citenamefont {Meng}, \citenamefont {Scherer}, \citenamefont {Uebelacker},
  \citenamefont {Assaad}, \citenamefont {Muramatsu}, \citenamefont
  {Honerkamp},\ and\ \citenamefont {Wessel}}]{PhysRevLett.109.126402}%
  \BibitemOpen
  \bibfield  {author} {\bibinfo {author} {\bibfnamefont {T.~C.}\ \bibnamefont
  {Lang}}, \bibinfo {author} {\bibfnamefont {Z.~Y.}\ \bibnamefont {Meng}},
  \bibinfo {author} {\bibfnamefont {M.~M.}\ \bibnamefont {Scherer}}, \bibinfo
  {author} {\bibfnamefont {S.}~\bibnamefont {Uebelacker}}, \bibinfo {author}
  {\bibfnamefont {F.~F.}\ \bibnamefont {Assaad}}, \bibinfo {author}
  {\bibfnamefont {A.}~\bibnamefont {Muramatsu}}, \bibinfo {author}
  {\bibfnamefont {C.}~\bibnamefont {Honerkamp}},\ and\ \bibinfo {author}
  {\bibfnamefont {S.}~\bibnamefont {Wessel}},\ }\bibfield  {title} {\bibinfo
  {title} {Antiferromagnetism in the hubbard model on the bernal-stacked
  honeycomb bilayer},\ }\href {https://doi.org/10.1103/PhysRevLett.109.126402}
  {\bibfield  {journal} {\bibinfo  {journal} {Phys. Rev. Lett.}\ }\textbf
  {\bibinfo {volume} {109}},\ \bibinfo {pages} {126402} (\bibinfo {year}
  {2012})}\BibitemShut {NoStop}%
\bibitem [{\citenamefont {Zhang}(2015)}]{ZHANG20159}%
  \BibitemOpen
  \bibfield  {author} {\bibinfo {author} {\bibfnamefont {F.}~\bibnamefont
  {Zhang}},\ }\bibfield  {title} {\bibinfo {title} {Spontaneous chiral symmetry
  breaking in bilayer graphene},\ }\href
  {https://doi.org/https://doi.org/10.1016/j.synthmet.2015.07.028} {\bibfield
  {journal} {\bibinfo  {journal} {Synthetic Metals}\ }\textbf {\bibinfo
  {volume} {210}},\ \bibinfo {pages} {9} (\bibinfo {year} {2015})},\ \bibinfo
  {note} {reviews of Current Advances in Graphene Science and
  Technology}\BibitemShut {NoStop}%
\bibitem [{\citenamefont {Raczkowski}\ \emph {et~al.}(2020)\citenamefont
  {Raczkowski}, \citenamefont {Peters}, \citenamefont {Ph\`ung}, \citenamefont
  {Takemori}, \citenamefont {Assaad}, \citenamefont {Honecker},\ and\
  \citenamefont {Vahedi}}]{PhysRevB.101.125103}%
  \BibitemOpen
  \bibfield  {author} {\bibinfo {author} {\bibfnamefont {M.}~\bibnamefont
  {Raczkowski}}, \bibinfo {author} {\bibfnamefont {R.}~\bibnamefont {Peters}},
  \bibinfo {author} {\bibfnamefont {T.~T.}\ \bibnamefont {Ph\`ung}}, \bibinfo
  {author} {\bibfnamefont {N.}~\bibnamefont {Takemori}}, \bibinfo {author}
  {\bibfnamefont {F.~F.}\ \bibnamefont {Assaad}}, \bibinfo {author}
  {\bibfnamefont {A.}~\bibnamefont {Honecker}},\ and\ \bibinfo {author}
  {\bibfnamefont {J.}~\bibnamefont {Vahedi}},\ }\bibfield  {title} {\bibinfo
  {title} {Hubbard model on the honeycomb lattice: From static and dynamical
  mean-field theories to lattice quantum monte carlo simulations},\ }\href
  {https://doi.org/10.1103/PhysRevB.101.125103} {\bibfield  {journal} {\bibinfo
   {journal} {Phys. Rev. B}\ }\textbf {\bibinfo {volume} {101}},\ \bibinfo
  {pages} {125103} (\bibinfo {year} {2020})}\BibitemShut {NoStop}%
\bibitem [{\citenamefont {Feldner}\ \emph {et~al.}(2010)\citenamefont
  {Feldner}, \citenamefont {Meng}, \citenamefont {Honecker}, \citenamefont
  {Cabra}, \citenamefont {Wessel},\ and\ \citenamefont
  {Assaad}}]{PhysRevB.81.115416}%
  \BibitemOpen
  \bibfield  {author} {\bibinfo {author} {\bibfnamefont {H.}~\bibnamefont
  {Feldner}}, \bibinfo {author} {\bibfnamefont {Z.~Y.}\ \bibnamefont {Meng}},
  \bibinfo {author} {\bibfnamefont {A.}~\bibnamefont {Honecker}}, \bibinfo
  {author} {\bibfnamefont {D.}~\bibnamefont {Cabra}}, \bibinfo {author}
  {\bibfnamefont {S.}~\bibnamefont {Wessel}},\ and\ \bibinfo {author}
  {\bibfnamefont {F.~F.}\ \bibnamefont {Assaad}},\ }\bibfield  {title}
  {\bibinfo {title} {Magnetism of finite graphene samples: Mean-field theory
  compared with exact diagonalization and quantum monte carlo simulations},\
  }\href {https://doi.org/10.1103/PhysRevB.81.115416} {\bibfield  {journal}
  {\bibinfo  {journal} {Phys. Rev. B}\ }\textbf {\bibinfo {volume} {81}},\
  \bibinfo {pages} {115416} (\bibinfo {year} {2010})}\BibitemShut {NoStop}%
\bibitem [{\citenamefont {Sorella}\ and\ \citenamefont
  {Tosatti}(1992)}]{S.Sorella_1992}%
  \BibitemOpen
  \bibfield  {author} {\bibinfo {author} {\bibfnamefont {S.}~\bibnamefont
  {Sorella}}\ and\ \bibinfo {author} {\bibfnamefont {E.}~\bibnamefont
  {Tosatti}},\ }\bibfield  {title} {\bibinfo {title} {Semi-metal-insulator
  transition of the hubbard model in the honeycomb lattice},\ }\href
  {https://doi.org/10.1209/0295-5075/19/8/007} {\bibfield  {journal} {\bibinfo
  {journal} {Europhysics Letters}\ }\textbf {\bibinfo {volume} {19}},\ \bibinfo
  {pages} {699} (\bibinfo {year} {1992})}\BibitemShut {NoStop}%
\bibitem [{\citenamefont {Pujari}\ \emph {et~al.}(2016)\citenamefont {Pujari},
  \citenamefont {Lang}, \citenamefont {Murthy},\ and\ \citenamefont
  {Kaul}}]{PhysRevLett.117.086404}%
  \BibitemOpen
  \bibfield  {author} {\bibinfo {author} {\bibfnamefont {S.}~\bibnamefont
  {Pujari}}, \bibinfo {author} {\bibfnamefont {T.~C.}\ \bibnamefont {Lang}},
  \bibinfo {author} {\bibfnamefont {G.}~\bibnamefont {Murthy}},\ and\ \bibinfo
  {author} {\bibfnamefont {R.~K.}\ \bibnamefont {Kaul}},\ }\bibfield  {title}
  {\bibinfo {title} {Interaction-induced dirac fermions from quadratic band
  touching in bilayer graphene},\ }\href
  {https://doi.org/10.1103/PhysRevLett.117.086404} {\bibfield  {journal}
  {\bibinfo  {journal} {Phys. Rev. Lett.}\ }\textbf {\bibinfo {volume} {117}},\
  \bibinfo {pages} {086404} (\bibinfo {year} {2016})}\BibitemShut {NoStop}%
\bibitem [{\citenamefont {Nandkishore}\ \emph
  {et~al.}(2012{\natexlab{a}})\citenamefont {Nandkishore}, \citenamefont
  {Levitov},\ and\ \citenamefont {Chubukov}}]{nandkishore2012chiral}%
  \BibitemOpen
  \bibfield  {author} {\bibinfo {author} {\bibfnamefont {R.}~\bibnamefont
  {Nandkishore}}, \bibinfo {author} {\bibfnamefont {L.~S.}\ \bibnamefont
  {Levitov}},\ and\ \bibinfo {author} {\bibfnamefont {A.~V.}\ \bibnamefont
  {Chubukov}},\ }\bibfield  {title} {\bibinfo {title} {Chiral superconductivity
  from repulsive interactions in doped graphene},\ }\href
  {https://doi.org/10.1038/nphys2208} {\bibfield  {journal} {\bibinfo
  {journal} {Nature Physics}\ }\textbf {\bibinfo {volume} {8}},\ \bibinfo
  {pages} {158} (\bibinfo {year} {2012}{\natexlab{a}})}\BibitemShut {NoStop}%
\bibitem [{\citenamefont {Nandkishore}\ and\ \citenamefont
  {Chubukov}(2012)}]{PhysRevB.86.115426}%
  \BibitemOpen
  \bibfield  {author} {\bibinfo {author} {\bibfnamefont {R.}~\bibnamefont
  {Nandkishore}}\ and\ \bibinfo {author} {\bibfnamefont {A.~V.}\ \bibnamefont
  {Chubukov}},\ }\bibfield  {title} {\bibinfo {title} {Interplay of
  superconductivity and spin-density-wave order in doped graphene},\ }\href
  {https://doi.org/10.1103/PhysRevB.86.115426} {\bibfield  {journal} {\bibinfo
  {journal} {Phys. Rev. B}\ }\textbf {\bibinfo {volume} {86}},\ \bibinfo
  {pages} {115426} (\bibinfo {year} {2012})}\BibitemShut {NoStop}%
\bibitem [{\citenamefont {Nandkishore}\ \emph {et~al.}(2014)\citenamefont
  {Nandkishore}, \citenamefont {Thomale},\ and\ \citenamefont
  {Chubukov}}]{PhysRevB.89.144501}%
  \BibitemOpen
  \bibfield  {author} {\bibinfo {author} {\bibfnamefont {R.}~\bibnamefont
  {Nandkishore}}, \bibinfo {author} {\bibfnamefont {R.}~\bibnamefont
  {Thomale}},\ and\ \bibinfo {author} {\bibfnamefont {A.~V.}\ \bibnamefont
  {Chubukov}},\ }\bibfield  {title} {\bibinfo {title} {Superconductivity from
  weak repulsion in hexagonal lattice systems},\ }\href
  {https://doi.org/10.1103/PhysRevB.89.144501} {\bibfield  {journal} {\bibinfo
  {journal} {Phys. Rev. B}\ }\textbf {\bibinfo {volume} {89}},\ \bibinfo
  {pages} {144501} (\bibinfo {year} {2014})}\BibitemShut {NoStop}%
\bibitem [{\citenamefont {Kiesel}\ \emph {et~al.}(2012)\citenamefont {Kiesel},
  \citenamefont {Platt}, \citenamefont {Hanke}, \citenamefont {Abanin},\ and\
  \citenamefont {Thomale}}]{PhysRevB.86.020507}%
  \BibitemOpen
  \bibfield  {author} {\bibinfo {author} {\bibfnamefont {M.~L.}\ \bibnamefont
  {Kiesel}}, \bibinfo {author} {\bibfnamefont {C.}~\bibnamefont {Platt}},
  \bibinfo {author} {\bibfnamefont {W.}~\bibnamefont {Hanke}}, \bibinfo
  {author} {\bibfnamefont {D.~A.}\ \bibnamefont {Abanin}},\ and\ \bibinfo
  {author} {\bibfnamefont {R.}~\bibnamefont {Thomale}},\ }\bibfield  {title}
  {\bibinfo {title} {Competing many-body instabilities and unconventional
  superconductivity in graphene},\ }\href
  {https://doi.org/10.1103/PhysRevB.86.020507} {\bibfield  {journal} {\bibinfo
  {journal} {Phys. Rev. B}\ }\textbf {\bibinfo {volume} {86}},\ \bibinfo
  {pages} {020507} (\bibinfo {year} {2012})}\BibitemShut {NoStop}%
\bibitem [{\citenamefont {Classen}\ \emph {et~al.}(2020)\citenamefont
  {Classen}, \citenamefont {Chubukov}, \citenamefont {Honerkamp},\ and\
  \citenamefont {Scherer}}]{PhysRevB.102.125141}%
  \BibitemOpen
  \bibfield  {author} {\bibinfo {author} {\bibfnamefont {L.}~\bibnamefont
  {Classen}}, \bibinfo {author} {\bibfnamefont {A.~V.}\ \bibnamefont
  {Chubukov}}, \bibinfo {author} {\bibfnamefont {C.}~\bibnamefont
  {Honerkamp}},\ and\ \bibinfo {author} {\bibfnamefont {M.~M.}\ \bibnamefont
  {Scherer}},\ }\bibfield  {title} {\bibinfo {title} {Competing orders at
  higher-order van hove points},\ }\href
  {https://doi.org/10.1103/PhysRevB.102.125141} {\bibfield  {journal} {\bibinfo
   {journal} {Phys. Rev. B}\ }\textbf {\bibinfo {volume} {102}},\ \bibinfo
  {pages} {125141} (\bibinfo {year} {2020})}\BibitemShut {NoStop}%
\bibitem [{\citenamefont {Li}(2012)}]{Li_2012}%
  \BibitemOpen
  \bibfield  {author} {\bibinfo {author} {\bibfnamefont {T.}~\bibnamefont
  {Li}},\ }\bibfield  {title} {\bibinfo {title} {Spontaneous quantum hall
  effect in quarter-doped hubbard model on honeycomb lattice and its possible
  realization in doped graphene system},\ }\href
  {https://doi.org/10.1209/0295-5075/97/37001} {\bibfield  {journal} {\bibinfo
  {journal} {Europhysics Letters}\ }\textbf {\bibinfo {volume} {97}},\ \bibinfo
  {pages} {37001} (\bibinfo {year} {2012})}\BibitemShut {NoStop}%
\bibitem [{\citenamefont {Martin}\ and\ \citenamefont
  {Batista}(2008)}]{PhysRevLett.101.156402}%
  \BibitemOpen
  \bibfield  {author} {\bibinfo {author} {\bibfnamefont {I.}~\bibnamefont
  {Martin}}\ and\ \bibinfo {author} {\bibfnamefont {C.~D.}\ \bibnamefont
  {Batista}},\ }\bibfield  {title} {\bibinfo {title} {Itinerant electron-driven
  chiral magnetic ordering and spontaneous quantum hall effect in triangular
  lattice models},\ }\href {https://doi.org/10.1103/PhysRevLett.101.156402}
  {\bibfield  {journal} {\bibinfo  {journal} {Phys. Rev. Lett.}\ }\textbf
  {\bibinfo {volume} {101}},\ \bibinfo {pages} {156402} (\bibinfo {year}
  {2008})}\BibitemShut {NoStop}%
\bibitem [{\citenamefont {Nandkishore}\ \emph
  {et~al.}(2012{\natexlab{b}})\citenamefont {Nandkishore}, \citenamefont
  {Chern},\ and\ \citenamefont {Chubukov}}]{Nandkishore2012}%
  \BibitemOpen
  \bibfield  {author} {\bibinfo {author} {\bibfnamefont {R.}~\bibnamefont
  {Nandkishore}}, \bibinfo {author} {\bibfnamefont {G.-W.}\ \bibnamefont
  {Chern}},\ and\ \bibinfo {author} {\bibfnamefont {A.~V.}\ \bibnamefont
  {Chubukov}},\ }\bibfield  {title} {\bibinfo {title} {Itinerant half-metal
  spin-density-wave state on the hexagonal lattice},\ }\href
  {https://doi.org/10.1103/PhysRevLett.108.227204} {\bibfield  {journal}
  {\bibinfo  {journal} {Phys. Rev. Lett.}\ }\textbf {\bibinfo {volume} {108}},\
  \bibinfo {pages} {227204} (\bibinfo {year} {2012}{\natexlab{b}})}\BibitemShut
  {NoStop}%
\bibitem [{\citenamefont {Kato}\ \emph {et~al.}(2010)\citenamefont {Kato},
  \citenamefont {Martin},\ and\ \citenamefont
  {Batista}}]{PhysRevLett.105.266405}%
  \BibitemOpen
  \bibfield  {author} {\bibinfo {author} {\bibfnamefont {Y.}~\bibnamefont
  {Kato}}, \bibinfo {author} {\bibfnamefont {I.}~\bibnamefont {Martin}},\ and\
  \bibinfo {author} {\bibfnamefont {C.~D.}\ \bibnamefont {Batista}},\
  }\bibfield  {title} {\bibinfo {title} {Stability of the spontaneous quantum
  hall state in the triangular kondo-lattice model},\ }\href
  {https://doi.org/10.1103/PhysRevLett.105.266405} {\bibfield  {journal}
  {\bibinfo  {journal} {Phys. Rev. Lett.}\ }\textbf {\bibinfo {volume} {105}},\
  \bibinfo {pages} {266405} (\bibinfo {year} {2010})}\BibitemShut {NoStop}%
\bibitem [{\citenamefont {Zhang}\ \emph {et~al.}(2011)\citenamefont {Zhang},
  \citenamefont {Jung}, \citenamefont {Fiete}, \citenamefont {Niu},\ and\
  \citenamefont {MacDonald}}]{PhysRevLett.106.156801}%
  \BibitemOpen
  \bibfield  {author} {\bibinfo {author} {\bibfnamefont {F.}~\bibnamefont
  {Zhang}}, \bibinfo {author} {\bibfnamefont {J.}~\bibnamefont {Jung}},
  \bibinfo {author} {\bibfnamefont {G.~A.}\ \bibnamefont {Fiete}}, \bibinfo
  {author} {\bibfnamefont {Q.}~\bibnamefont {Niu}},\ and\ \bibinfo {author}
  {\bibfnamefont {A.~H.}\ \bibnamefont {MacDonald}},\ }\bibfield  {title}
  {\bibinfo {title} {Spontaneous quantum hall states in chirally stacked
  few-layer graphene systems},\ }\href
  {https://doi.org/10.1103/PhysRevLett.106.156801} {\bibfield  {journal}
  {\bibinfo  {journal} {Phys. Rev. Lett.}\ }\textbf {\bibinfo {volume} {106}},\
  \bibinfo {pages} {156801} (\bibinfo {year} {2011})}\BibitemShut {NoStop}%
\bibitem [{\citenamefont {Weitz}\ \emph {et~al.}(2010)\citenamefont {Weitz},
  \citenamefont {Allen}, \citenamefont {Feldman}, \citenamefont {Martin},\ and\
  \citenamefont {Yacoby}}]{doi:10.1126/science.1194988}%
  \BibitemOpen
  \bibfield  {author} {\bibinfo {author} {\bibfnamefont {R.~T.}\ \bibnamefont
  {Weitz}}, \bibinfo {author} {\bibfnamefont {M.~T.}\ \bibnamefont {Allen}},
  \bibinfo {author} {\bibfnamefont {B.~E.}\ \bibnamefont {Feldman}}, \bibinfo
  {author} {\bibfnamefont {J.}~\bibnamefont {Martin}},\ and\ \bibinfo {author}
  {\bibfnamefont {A.}~\bibnamefont {Yacoby}},\ }\bibfield  {title} {\bibinfo
  {title} {Broken-symmetry states in doubly gated suspended bilayer graphene},\
  }\href {https://doi.org/10.1126/science.1194988} {\bibfield  {journal}
  {\bibinfo  {journal} {Science}\ }\textbf {\bibinfo {volume} {330}},\ \bibinfo
  {pages} {812} (\bibinfo {year} {2010})},\ \Eprint
  {https://arxiv.org/abs/https://www.science.org/doi/pdf/10.1126/science.1194988}
  {https://www.science.org/doi/pdf/10.1126/science.1194988} \BibitemShut
  {NoStop}%
\bibitem [{\citenamefont {Velasco}\ \emph {et~al.}(2012)\citenamefont
  {Velasco}, \citenamefont {Jing}, \citenamefont {Bao}, \citenamefont {Lee},
  \citenamefont {Kratz}, \citenamefont {Aji}, \citenamefont {Bockrath},
  \citenamefont {Lau}, \citenamefont {Varma}, \citenamefont {Stillwell},
  \citenamefont {Smirnov}, \citenamefont {Zhang}, \citenamefont {Jung},\ and\
  \citenamefont {MacDonald}}]{Velasco2012}%
  \BibitemOpen
  \bibfield  {author} {\bibinfo {author} {\bibfnamefont {J.}~\bibnamefont
  {Velasco}}, \bibinfo {author} {\bibfnamefont {L.}~\bibnamefont {Jing}},
  \bibinfo {author} {\bibfnamefont {W.}~\bibnamefont {Bao}}, \bibinfo {author}
  {\bibfnamefont {Y.}~\bibnamefont {Lee}}, \bibinfo {author} {\bibfnamefont
  {P.}~\bibnamefont {Kratz}}, \bibinfo {author} {\bibfnamefont
  {V.}~\bibnamefont {Aji}}, \bibinfo {author} {\bibfnamefont {M.}~\bibnamefont
  {Bockrath}}, \bibinfo {author} {\bibfnamefont {C.~N.}\ \bibnamefont {Lau}},
  \bibinfo {author} {\bibfnamefont {C.}~\bibnamefont {Varma}}, \bibinfo
  {author} {\bibfnamefont {R.}~\bibnamefont {Stillwell}}, \bibinfo {author}
  {\bibfnamefont {D.}~\bibnamefont {Smirnov}}, \bibinfo {author} {\bibfnamefont
  {F.}~\bibnamefont {Zhang}}, \bibinfo {author} {\bibfnamefont
  {J.}~\bibnamefont {Jung}},\ and\ \bibinfo {author} {\bibfnamefont {A.~H.}\
  \bibnamefont {MacDonald}},\ }\bibfield  {title} {\bibinfo {title} {Transport
  spectroscopy of symmetry-broken insulating states in bilayer graphene},\
  }\href {https://doi.org/10.1038/nnano.2011.251} {\bibfield  {journal}
  {\bibinfo  {journal} {Nature Nanotechnology}\ }\textbf {\bibinfo {volume}
  {7}},\ \bibinfo {pages} {156} (\bibinfo {year} {2012})}\BibitemShut {NoStop}%
\bibitem [{\citenamefont {Freitag}\ \emph {et~al.}(2012)\citenamefont
  {Freitag}, \citenamefont {Trbovic}, \citenamefont {Weiss},\ and\
  \citenamefont {Sch\"onenberger}}]{PhysRevLett.108.076602}%
  \BibitemOpen
  \bibfield  {author} {\bibinfo {author} {\bibfnamefont {F.}~\bibnamefont
  {Freitag}}, \bibinfo {author} {\bibfnamefont {J.}~\bibnamefont {Trbovic}},
  \bibinfo {author} {\bibfnamefont {M.}~\bibnamefont {Weiss}},\ and\ \bibinfo
  {author} {\bibfnamefont {C.}~\bibnamefont {Sch\"onenberger}},\ }\bibfield
  {title} {\bibinfo {title} {Spontaneously gapped ground state in suspended
  bilayer graphene},\ }\href {https://doi.org/10.1103/PhysRevLett.108.076602}
  {\bibfield  {journal} {\bibinfo  {journal} {Phys. Rev. Lett.}\ }\textbf
  {\bibinfo {volume} {108}},\ \bibinfo {pages} {076602} (\bibinfo {year}
  {2012})}\BibitemShut {NoStop}%
\bibitem [{\citenamefont {Martin}\ \emph {et~al.}(2010)\citenamefont {Martin},
  \citenamefont {Feldman}, \citenamefont {Weitz}, \citenamefont {Allen},\ and\
  \citenamefont {Yacoby}}]{PhysRevLett.105.256806}%
  \BibitemOpen
  \bibfield  {author} {\bibinfo {author} {\bibfnamefont {J.}~\bibnamefont
  {Martin}}, \bibinfo {author} {\bibfnamefont {B.~E.}\ \bibnamefont {Feldman}},
  \bibinfo {author} {\bibfnamefont {R.~T.}\ \bibnamefont {Weitz}}, \bibinfo
  {author} {\bibfnamefont {M.~T.}\ \bibnamefont {Allen}},\ and\ \bibinfo
  {author} {\bibfnamefont {A.}~\bibnamefont {Yacoby}},\ }\bibfield  {title}
  {\bibinfo {title} {Local compressibility measurements of correlated states in
  suspended bilayer graphene},\ }\href
  {https://doi.org/10.1103/PhysRevLett.105.256806} {\bibfield  {journal}
  {\bibinfo  {journal} {Phys. Rev. Lett.}\ }\textbf {\bibinfo {volume} {105}},\
  \bibinfo {pages} {256806} (\bibinfo {year} {2010})}\BibitemShut {NoStop}%
\bibitem [{\citenamefont {Ki}\ and\ \citenamefont {Morpurgo}(2013)}]{Ki2013}%
  \BibitemOpen
  \bibfield  {author} {\bibinfo {author} {\bibfnamefont {D.-K.}\ \bibnamefont
  {Ki}}\ and\ \bibinfo {author} {\bibfnamefont {A.~F.}\ \bibnamefont
  {Morpurgo}},\ }\bibfield  {title} {\bibinfo {title} {High-quality
  multiterminal suspended graphene devices},\ }\href
  {https://doi.org/10.1021/nl402462q} {\bibfield  {journal} {\bibinfo
  {journal} {Nano Letters}\ }\textbf {\bibinfo {volume} {13}},\ \bibinfo
  {pages} {5165} (\bibinfo {year} {2013})}\BibitemShut {NoStop}%
\bibitem [{\citenamefont {Mayorov}\ \emph {et~al.}(2011)\citenamefont
  {Mayorov}, \citenamefont {Elias}, \citenamefont {Mucha-Kruczynski},
  \citenamefont {Gorbachev}, \citenamefont {Tudorovskiy}, \citenamefont
  {Zhukov}, \citenamefont {Morozov}, \citenamefont {Katsnelson}, \citenamefont
  {Fal’ko}, \citenamefont {Geim},\ and\ \citenamefont
  {Novoselov}}]{doi:10.1126/science.1208683}%
  \BibitemOpen
  \bibfield  {author} {\bibinfo {author} {\bibfnamefont {A.~S.}\ \bibnamefont
  {Mayorov}}, \bibinfo {author} {\bibfnamefont {D.~C.}\ \bibnamefont {Elias}},
  \bibinfo {author} {\bibfnamefont {M.}~\bibnamefont {Mucha-Kruczynski}},
  \bibinfo {author} {\bibfnamefont {R.~V.}\ \bibnamefont {Gorbachev}}, \bibinfo
  {author} {\bibfnamefont {T.}~\bibnamefont {Tudorovskiy}}, \bibinfo {author}
  {\bibfnamefont {A.}~\bibnamefont {Zhukov}}, \bibinfo {author} {\bibfnamefont
  {S.~V.}\ \bibnamefont {Morozov}}, \bibinfo {author} {\bibfnamefont {M.~I.}\
  \bibnamefont {Katsnelson}}, \bibinfo {author} {\bibfnamefont {V.~I.}\
  \bibnamefont {Fal’ko}}, \bibinfo {author} {\bibfnamefont {A.~K.}\
  \bibnamefont {Geim}},\ and\ \bibinfo {author} {\bibfnamefont {K.~S.}\
  \bibnamefont {Novoselov}},\ }\bibfield  {title} {\bibinfo {title}
  {Interaction-driven spectrum reconstruction in bilayer graphene},\ }\href
  {https://doi.org/10.1126/science.1208683} {\bibfield  {journal} {\bibinfo
  {journal} {Science}\ }\textbf {\bibinfo {volume} {333}},\ \bibinfo {pages}
  {860} (\bibinfo {year} {2011})},\ \Eprint
  {https://arxiv.org/abs/https://www.science.org/doi/pdf/10.1126/science.1208683}
  {https://www.science.org/doi/pdf/10.1126/science.1208683} \BibitemShut
  {NoStop}%
\bibitem [{\citenamefont {Feldman}\ \emph {et~al.}(2009)\citenamefont
  {Feldman}, \citenamefont {Martin},\ and\ \citenamefont
  {Yacoby}}]{Feldman2009}%
  \BibitemOpen
  \bibfield  {author} {\bibinfo {author} {\bibfnamefont {B.~E.}\ \bibnamefont
  {Feldman}}, \bibinfo {author} {\bibfnamefont {J.}~\bibnamefont {Martin}},\
  and\ \bibinfo {author} {\bibfnamefont {A.}~\bibnamefont {Yacoby}},\
  }\bibfield  {title} {\bibinfo {title} {Broken-symmetry states and divergent
  resistance in suspended bilayer graphene},\ }\href
  {https://doi.org/10.1038/nphys1406} {\bibfield  {journal} {\bibinfo
  {journal} {Nature Physics}\ }\textbf {\bibinfo {volume} {5}},\ \bibinfo
  {pages} {889} (\bibinfo {year} {2009})}\BibitemShut {NoStop}%
\bibitem [{\citenamefont {Geisenhof}\ \emph {et~al.}(2021)\citenamefont
  {Geisenhof}, \citenamefont {Winterer}, \citenamefont {Seiler}, \citenamefont
  {Lenz}, \citenamefont {Xu}, \citenamefont {Zhang},\ and\ \citenamefont
  {Weitz}}]{Geisenhof2021}%
  \BibitemOpen
  \bibfield  {author} {\bibinfo {author} {\bibfnamefont {F.~R.}\ \bibnamefont
  {Geisenhof}}, \bibinfo {author} {\bibfnamefont {F.}~\bibnamefont {Winterer}},
  \bibinfo {author} {\bibfnamefont {A.~M.}\ \bibnamefont {Seiler}}, \bibinfo
  {author} {\bibfnamefont {J.}~\bibnamefont {Lenz}}, \bibinfo {author}
  {\bibfnamefont {T.}~\bibnamefont {Xu}}, \bibinfo {author} {\bibfnamefont
  {F.}~\bibnamefont {Zhang}},\ and\ \bibinfo {author} {\bibfnamefont {R.~T.}\
  \bibnamefont {Weitz}},\ }\bibfield  {title} {\bibinfo {title} {Quantum
  anomalous hall octet driven by orbital magnetism in bilayer graphene},\
  }\href {https://doi.org/10.1038/s41586-021-03849-w} {\bibfield  {journal}
  {\bibinfo  {journal} {Nature}\ }\textbf {\bibinfo {volume} {598}},\ \bibinfo
  {pages} {53} (\bibinfo {year} {2021})}\BibitemShut {NoStop}%
\bibitem [{\citenamefont {Yang}\ \emph {et~al.}(2022)\citenamefont {Yang},
  \citenamefont {Fecher}, \citenamefont {Wang}, \citenamefont {Kühne},\ and\
  \citenamefont {Smet}}]{Yang_2022}%
  \BibitemOpen
  \bibfield  {author} {\bibinfo {author} {\bibfnamefont {S.}~\bibnamefont
  {Yang}}, \bibinfo {author} {\bibfnamefont {S.}~\bibnamefont {Fecher}},
  \bibinfo {author} {\bibfnamefont {Q.}~\bibnamefont {Wang}}, \bibinfo {author}
  {\bibfnamefont {M.}~\bibnamefont {Kühne}},\ and\ \bibinfo {author}
  {\bibfnamefont {J.~H.}\ \bibnamefont {Smet}},\ }\bibfield  {title} {\bibinfo
  {title} {Device level reversible potassium intercalation into bilayer
  graphene},\ }\href {https://doi.org/10.1088/2053-1583/ac58a1} {\bibfield
  {journal} {\bibinfo  {journal} {2D Materials}\ }\textbf {\bibinfo {volume}
  {9}},\ \bibinfo {pages} {025020} (\bibinfo {year} {2022})}\BibitemShut
  {NoStop}%
\bibitem [{\citenamefont {Link}\ \emph {et~al.}(2019)\citenamefont {Link},
  \citenamefont {Forti}, \citenamefont {St\"ohr}, \citenamefont {K\"uster},
  \citenamefont {R\"osner}, \citenamefont {Hirschmeier}, \citenamefont {Chen},
  \citenamefont {Avila}, \citenamefont {Asensio}, \citenamefont {Zakharov},
  \citenamefont {Wehling}, \citenamefont {Lichtenstein}, \citenamefont
  {Katsnelson},\ and\ \citenamefont {Starke}}]{PhysRevB.100.121407}%
  \BibitemOpen
  \bibfield  {author} {\bibinfo {author} {\bibfnamefont {S.}~\bibnamefont
  {Link}}, \bibinfo {author} {\bibfnamefont {S.}~\bibnamefont {Forti}},
  \bibinfo {author} {\bibfnamefont {A.}~\bibnamefont {St\"ohr}}, \bibinfo
  {author} {\bibfnamefont {K.}~\bibnamefont {K\"uster}}, \bibinfo {author}
  {\bibfnamefont {M.}~\bibnamefont {R\"osner}}, \bibinfo {author}
  {\bibfnamefont {D.}~\bibnamefont {Hirschmeier}}, \bibinfo {author}
  {\bibfnamefont {C.}~\bibnamefont {Chen}}, \bibinfo {author} {\bibfnamefont
  {J.}~\bibnamefont {Avila}}, \bibinfo {author} {\bibfnamefont {M.~C.}\
  \bibnamefont {Asensio}}, \bibinfo {author} {\bibfnamefont {A.~A.}\
  \bibnamefont {Zakharov}}, \bibinfo {author} {\bibfnamefont {T.~O.}\
  \bibnamefont {Wehling}}, \bibinfo {author} {\bibfnamefont {A.~I.}\
  \bibnamefont {Lichtenstein}}, \bibinfo {author} {\bibfnamefont {M.~I.}\
  \bibnamefont {Katsnelson}},\ and\ \bibinfo {author} {\bibfnamefont
  {U.}~\bibnamefont {Starke}},\ }\bibfield  {title} {\bibinfo {title}
  {Introducing strong correlation effects into graphene by gadolinium
  intercalation},\ }\href {https://doi.org/10.1103/PhysRevB.100.121407}
  {\bibfield  {journal} {\bibinfo  {journal} {Phys. Rev. B}\ }\textbf {\bibinfo
  {volume} {100}},\ \bibinfo {pages} {121407} (\bibinfo {year}
  {2019})}\BibitemShut {NoStop}%
\bibitem [{\citenamefont {Rosenzweig}\ \emph {et~al.}(2020)\citenamefont
  {Rosenzweig}, \citenamefont {Karakachian}, \citenamefont {Marchenko},
  \citenamefont {K\"uster},\ and\ \citenamefont
  {Starke}}]{PhysRevLett.125.176403}%
  \BibitemOpen
  \bibfield  {author} {\bibinfo {author} {\bibfnamefont {P.}~\bibnamefont
  {Rosenzweig}}, \bibinfo {author} {\bibfnamefont {H.}~\bibnamefont
  {Karakachian}}, \bibinfo {author} {\bibfnamefont {D.}~\bibnamefont
  {Marchenko}}, \bibinfo {author} {\bibfnamefont {K.}~\bibnamefont
  {K\"uster}},\ and\ \bibinfo {author} {\bibfnamefont {U.}~\bibnamefont
  {Starke}},\ }\bibfield  {title} {\bibinfo {title} {Overdoping graphene beyond
  the van hove singularity},\ }\href
  {https://doi.org/10.1103/PhysRevLett.125.176403} {\bibfield  {journal}
  {\bibinfo  {journal} {Phys. Rev. Lett.}\ }\textbf {\bibinfo {volume} {125}},\
  \bibinfo {pages} {176403} (\bibinfo {year} {2020})}\BibitemShut {NoStop}%
\bibitem [{\citenamefont {Link}(2017)}]{PhDLink}%
  \BibitemOpen
  \bibfield  {author} {\bibinfo {author} {\bibfnamefont {S.}~\bibnamefont
  {Link}},\ }\href
  {https://open.fau.de/items/aa77adc9-f2a0-4824-aee1-16888921e1c4} {\bibinfo
  {title} {Intercalation of graphene on sic(0001): Ultra-high doping levels and
  new 2d materials}} (\bibinfo {year} {2017})\BibitemShut {NoStop}%
\bibitem [{\citenamefont {McChesney}\ \emph {et~al.}(2010)\citenamefont
  {McChesney}, \citenamefont {Bostwick}, \citenamefont {Ohta}, \citenamefont
  {Seyller}, \citenamefont {Horn}, \citenamefont {Gonz\'alez},\ and\
  \citenamefont {Rotenberg}}]{PhysRevLett.104.136803}%
  \BibitemOpen
  \bibfield  {author} {\bibinfo {author} {\bibfnamefont {J.~L.}\ \bibnamefont
  {McChesney}}, \bibinfo {author} {\bibfnamefont {A.}~\bibnamefont {Bostwick}},
  \bibinfo {author} {\bibfnamefont {T.}~\bibnamefont {Ohta}}, \bibinfo {author}
  {\bibfnamefont {T.}~\bibnamefont {Seyller}}, \bibinfo {author} {\bibfnamefont
  {K.}~\bibnamefont {Horn}}, \bibinfo {author} {\bibfnamefont {J.}~\bibnamefont
  {Gonz\'alez}},\ and\ \bibinfo {author} {\bibfnamefont {E.}~\bibnamefont
  {Rotenberg}},\ }\bibfield  {title} {\bibinfo {title} {Extended van hove
  singularity and superconducting instability in doped graphene},\ }\href
  {https://doi.org/10.1103/PhysRevLett.104.136803} {\bibfield  {journal}
  {\bibinfo  {journal} {Phys. Rev. Lett.}\ }\textbf {\bibinfo {volume} {104}},\
  \bibinfo {pages} {136803} (\bibinfo {year} {2010})}\BibitemShut {NoStop}%
\bibitem [{\citenamefont {Zhou}\ \emph {et~al.}(2022)\citenamefont {Zhou},
  \citenamefont {Holleis}, \citenamefont {Saito}, \citenamefont {Cohen},
  \citenamefont {Huynh}, \citenamefont {Patterson}, \citenamefont {Yang},
  \citenamefont {Taniguchi}, \citenamefont {Watanabe},\ and\ \citenamefont
  {Young}}]{doi:10.1126/science.abm8386}%
  \BibitemOpen
  \bibfield  {author} {\bibinfo {author} {\bibfnamefont {H.}~\bibnamefont
  {Zhou}}, \bibinfo {author} {\bibfnamefont {L.}~\bibnamefont {Holleis}},
  \bibinfo {author} {\bibfnamefont {Y.}~\bibnamefont {Saito}}, \bibinfo
  {author} {\bibfnamefont {L.}~\bibnamefont {Cohen}}, \bibinfo {author}
  {\bibfnamefont {W.}~\bibnamefont {Huynh}}, \bibinfo {author} {\bibfnamefont
  {C.~L.}\ \bibnamefont {Patterson}}, \bibinfo {author} {\bibfnamefont
  {F.}~\bibnamefont {Yang}}, \bibinfo {author} {\bibfnamefont {T.}~\bibnamefont
  {Taniguchi}}, \bibinfo {author} {\bibfnamefont {K.}~\bibnamefont
  {Watanabe}},\ and\ \bibinfo {author} {\bibfnamefont {A.~F.}\ \bibnamefont
  {Young}},\ }\bibfield  {title} {\bibinfo {title} {Isospin magnetism and
  spin-polarized superconductivity in bernal bilayer graphene},\ }\href
  {https://doi.org/10.1126/science.abm8386} {\bibfield  {journal} {\bibinfo
  {journal} {Science}\ }\textbf {\bibinfo {volume} {375}},\ \bibinfo {pages}
  {774} (\bibinfo {year} {2022})},\ \Eprint
  {https://arxiv.org/abs/https://www.science.org/doi/pdf/10.1126/science.abm8386}
  {https://www.science.org/doi/pdf/10.1126/science.abm8386} \BibitemShut
  {NoStop}%
\bibitem [{\citenamefont {Seiler}\ \emph {et~al.}(2022)\citenamefont {Seiler},
  \citenamefont {Geisenhof}, \citenamefont {Winterer}, \citenamefont
  {Watanabe}, \citenamefont {Taniguchi}, \citenamefont {Xu}, \citenamefont
  {Zhang},\ and\ \citenamefont {Weitz}}]{Seiler2022}%
  \BibitemOpen
  \bibfield  {author} {\bibinfo {author} {\bibfnamefont {A.~M.}\ \bibnamefont
  {Seiler}}, \bibinfo {author} {\bibfnamefont {F.~R.}\ \bibnamefont
  {Geisenhof}}, \bibinfo {author} {\bibfnamefont {F.}~\bibnamefont {Winterer}},
  \bibinfo {author} {\bibfnamefont {K.}~\bibnamefont {Watanabe}}, \bibinfo
  {author} {\bibfnamefont {T.}~\bibnamefont {Taniguchi}}, \bibinfo {author}
  {\bibfnamefont {T.}~\bibnamefont {Xu}}, \bibinfo {author} {\bibfnamefont
  {F.}~\bibnamefont {Zhang}},\ and\ \bibinfo {author} {\bibfnamefont {R.~T.}\
  \bibnamefont {Weitz}},\ }\bibfield  {title} {\bibinfo {title} {Quantum
  cascade of correlated phases in trigonally warped bilayer graphene},\ }\href
  {https://doi.org/10.1038/s41586-022-04937-1} {\bibfield  {journal} {\bibinfo
  {journal} {Nature}\ }\textbf {\bibinfo {volume} {608}},\ \bibinfo {pages}
  {298} (\bibinfo {year} {2022})}\BibitemShut {NoStop}%
\bibitem [{\citenamefont {Seiler}\ \emph {et~al.}(2024)\citenamefont {Seiler},
  \citenamefont {Statz}, \citenamefont {Weimer}, \citenamefont {Jacobsen},
  \citenamefont {Watanabe}, \citenamefont {Taniguchi}, \citenamefont {Dong},
  \citenamefont {Levitov},\ and\ \citenamefont
  {Weitz}}]{PhysRevLett.133.066301}%
  \BibitemOpen
  \bibfield  {author} {\bibinfo {author} {\bibfnamefont {A.~M.}\ \bibnamefont
  {Seiler}}, \bibinfo {author} {\bibfnamefont {M.}~\bibnamefont {Statz}},
  \bibinfo {author} {\bibfnamefont {I.}~\bibnamefont {Weimer}}, \bibinfo
  {author} {\bibfnamefont {N.}~\bibnamefont {Jacobsen}}, \bibinfo {author}
  {\bibfnamefont {K.}~\bibnamefont {Watanabe}}, \bibinfo {author}
  {\bibfnamefont {T.}~\bibnamefont {Taniguchi}}, \bibinfo {author}
  {\bibfnamefont {Z.}~\bibnamefont {Dong}}, \bibinfo {author} {\bibfnamefont
  {L.~S.}\ \bibnamefont {Levitov}},\ and\ \bibinfo {author} {\bibfnamefont
  {R.~T.}\ \bibnamefont {Weitz}},\ }\bibfield  {title} {\bibinfo {title}
  {Interaction-driven quasi-insulating ground states of gapped electron-doped
  bilayer graphene},\ }\href {https://doi.org/10.1103/PhysRevLett.133.066301}
  {\bibfield  {journal} {\bibinfo  {journal} {Phys. Rev. Lett.}\ }\textbf
  {\bibinfo {volume} {133}},\ \bibinfo {pages} {066301} (\bibinfo {year}
  {2024})}\BibitemShut {NoStop}%
\bibitem [{\citenamefont {Sachdev}\ \emph {et~al.}(2019)\citenamefont
  {Sachdev}, \citenamefont {Scammell}, \citenamefont {Scheurer},\ and\
  \citenamefont {Tarnopolsky}}]{Sachdev2019}%
  \BibitemOpen
  \bibfield  {author} {\bibinfo {author} {\bibfnamefont {S.}~\bibnamefont
  {Sachdev}}, \bibinfo {author} {\bibfnamefont {H.~D.}\ \bibnamefont
  {Scammell}}, \bibinfo {author} {\bibfnamefont {M.~S.}\ \bibnamefont
  {Scheurer}},\ and\ \bibinfo {author} {\bibfnamefont {G.}~\bibnamefont
  {Tarnopolsky}},\ }\bibfield  {title} {\bibinfo {title} {Gauge theory for the
  cuprates near optimal doping},\ }\href
  {https://doi.org/10.1103/PhysRevB.99.054516} {\bibfield  {journal} {\bibinfo
  {journal} {Phys. Rev. B}\ }\textbf {\bibinfo {volume} {99}},\ \bibinfo
  {pages} {054516} (\bibinfo {year} {2019})}\BibitemShut {NoStop}%
\bibitem [{\citenamefont {Zachar}\ \emph {et~al.}(1998)\citenamefont {Zachar},
  \citenamefont {Kivelson},\ and\ \citenamefont {Emery}}]{ZacharKivelson1998}%
  \BibitemOpen
  \bibfield  {author} {\bibinfo {author} {\bibfnamefont {O.}~\bibnamefont
  {Zachar}}, \bibinfo {author} {\bibfnamefont {S.~A.}\ \bibnamefont
  {Kivelson}},\ and\ \bibinfo {author} {\bibfnamefont {V.~J.}\ \bibnamefont
  {Emery}},\ }\bibfield  {title} {\bibinfo {title} {{Landau theory of stripe
  phases in cuprates and nickelates}},\ }\href
  {https://doi.org/10.1103/PhysRevB.57.1422} {\bibfield  {journal} {\bibinfo
  {journal} {Phys. Rev. B}\ }\textbf {\bibinfo {volume} {57}},\ \bibinfo
  {pages} {1422} (\bibinfo {year} {1998})}\BibitemShut {NoStop}%
\bibitem [{\citenamefont {Classen}\ and\ \citenamefont
  {Betouras}(2024)}]{HOVHSreview}%
  \BibitemOpen
  \bibfield  {author} {\bibinfo {author} {\bibfnamefont {L.}~\bibnamefont
  {Classen}}\ and\ \bibinfo {author} {\bibfnamefont {J.~J.}\ \bibnamefont
  {Betouras}},\ }\bibfield  {title} {\bibinfo {title} {High-order van hove
  singularities and their connection to flat bands},\ }\bibfield  {journal}
  {\bibinfo  {journal} {Annual Review of Condensed Matter Physics}\ }\href
  {https://doi.org/https://doi.org/10.1146/annurev-conmatphys-042924-015000}
  {https://doi.org/10.1146/annurev-conmatphys-042924-015000} (\bibinfo {year}
  {2024})\BibitemShut {NoStop}%
\end{thebibliography}%

\end{document}